\documentclass[prd,twocolumn,english,preprintnumbers,amsmath,amssymb,nofootinbib,superscriptaddress,a4]{revtex4}

\usepackage[latin1]{inputenc}
\usepackage{graphicx}
\usepackage{color}
\usepackage{axodraw}

\usepackage{color}

\usepackage{bm}
\usepackage{bbm}
\usepackage{amsmath}

\usepackage{dsfont}

\newcommand{\be}{\begin{eqnarray}}
\newcommand{\ee}{\end{eqnarray}}

\newcommand{\nn}{\nonumber }

\newcommand{\yb}{\bar\psi}
\newcommand{\xsb}{$\chi$SB}
\newcommand{\Nf}{N_{\text{f}}}
\newcommand{\Nfcr}{N_{\text{f,cr}}}
\newcommand{\kcr}{k_{\text{cr}}}
\newcommand{\ksb}{k_{\text{SB}}}
\newcommand{\Nc}{N_{\text{c}}}
\newcommand{\pat}{\partial_t}
\newcommand{\Eqref}[1]{Eq.~\eqref{#1}}
\newcommand{\lF}{l_1^{\text{(F)},4}}
\newcommand{\lFB}{l^{\textrm{(FB)},4}_{1,2}}
\newcommand{\lFBo}{l^{\textrm{(FB)},4}_{1,1}}
\newcommand{\casel}[2]{{\scriptstyle \frac{#1}{#2}}}

\newcommand{\Cas}{C_2(\Nc)}

\newcommand{\LQCD}{\Lambda_{\text{QCD}}}
\newcommand{\MSbar}{\overline{\text{MS}}} 
\newcommand{\hath}[1]{#1}
\newcommand{\lp}{\hath{\lambda}_{+}}
\newcommand{\lm}{\hath{\lambda}_{-}}

\newcommand{\lsf}{\hath{\lambda}_{\sigma}}
\newcommand{\lva}{\hath{\lambda}_{\text{VA}}}

\def\slash#1{\setbox0=\hbox{$#1$}               % set a box for #1
   \dimen0=\wd0                                 % and get its size
   \setbox1=\hbox{/} \dimen1=\wd1               % get size of /
   \ifdim\dimen0>\dimen1                        % #1 is bigger
      \rlap{\hbox to \dimen0{\hfil/\hfil}}      % so center / in box
      #1                                        % and print #1
   \else                            
      \rlap{\hbox to \dimen1{\hfil$#1$\hfil}}   % so center #1
      /                                         % and print /
   \fi}                                         %

\usepackage{babel}
\makeatother

\begin{document}

\title{Beyond Miransky Scaling}
\author{Jens Braun}
\affiliation{Theoretisch-Physikalisches Institut, Friedrich-Schiller-Universit\"at Jena, Max-Wien-Platz 1, D-07743 Jena, Germany}
\author{Christian S. Fischer}
\affiliation{Institut f\"ur Theoretische Physik, Justus-Liebig-Universit\"at Giessen, {Heinrich-Buff-Ring 16, 35392 Giessen}, Germany}
\author{Holger Gies}
\affiliation{Theoretisch-Physikalisches Institut, Friedrich-Schiller-Universit\"at Jena, Max-Wien-Platz 1, D-07743 Jena, Germany}

\begin{abstract}
  We study the scaling behavior of physical observables in strongly-flavored
  asymptotically free gauge theories, such as many-flavor QCD. Such theories
  approach a quantum critical point when the number of fermion flavors is
  increased. It is well-known that physical observables at this quantum
  critical point exhibit an exponential scaling behavior (Miransky scaling),
  provided the gauge coupling is considered as a constant external
  parameter. This scaling behavior is modified when the scale dependence of
  the gauge coupling is taken into account. Provided that the gauge coupling
  approaches an IR fixed point, we derive the resulting {\it universal}
  power-law corrections to the exponential scaling behavior and show that they
  are uniquely determined by the IR critical exponent of the gauge
  coupling. To illustrate our findings, we compute the universal corrections
  in many-flavor QCD with the aid of nonperturbative functional
  renormalization group methods. In this case, we expect the power-law scaling
  to be quantitatively more relevant if the theories are probed, for instance,
  at integer $\Nf$ as done in lattice simulations.
\end{abstract}

\maketitle

\section{Introduction}
Strongly-flavored asymptotically free theories, such as QCD and QED${}_3$
are currently very actively researched. In particular, QCD with many quark flavors 
has drawn a lot of attention in recent years. 
On the one hand, the number of (massless) fermions can be considered as 
an external parameter. Such gauge theories are then expected to exhibit a 
quantum phase transition from a chirally broken to a conformal phase when the number of 
fermion flavors is increased. On the other hand, the understanding of strongly-flavored 
gauge theories underlies (walking) technicolor-like scenarios for the Higgs 
sector, see e.~g. Refs.~\cite{Weinberg:1979bn,Holdom:1981rm,Hong:2004td,Sannino:2004qp,Dietrich:2005jn,
Dietrich:2006cm,Ryttov:2007sr,Antipin:2009wr,Sannino:2009za}.

The phase structure of gauge theories with $\Nf$ fermions can indeed be rich,
as simple considerations may already suggest. Due to the screening property of
fermionic fluctuations, asymptotic freedom is lost for large $\Nf$. For
instance, SU($\Nc$) gauge theory with $\Nf$ fermions is no longer
asymptotically free (a.f.) for $\Nf>\Nf^{\text{a.f.}}:=\frac{11}{2}
\Nc$. Another special fermion number $\Nf^{\text{CBZ}}$ potentially
  exists denoting the minimum flavor number for the occurrence of an infrared
  fixed point  $g^2_{\ast}$ of the running gauge coupling. For instance, the
 two-loop $\beta$ function of the gauge coupling $g^2$ exhibits the so-called
 Caswell-Banks-Zaks (CBZ) fixed point \cite{Caswell:1974gg}, as the screening
 nature of fermion fluctuations dominates the two-loop coeefficient for
 $\Nf>\Nf^{\text{CBZ}}$. For instance for SU(3), 
 we have $\Nf^{\text{CBZ}}\simeq 8.05$ in the two-loop approximation. A
perturbative treatment of the theory seems possible near $\Nf^{\text{a.f.}}$,
$\Nf\lesssim\Nf^{\text{a.f.}}$, where $g^2_{\ast}$ is small, indicating the
existence of a conformally invariant limit in the deep
infrared~\cite{Banks:1981nn}. For decreasing $\Nf$, $g^2_{\ast}$ becomes
larger, suggesting the onset of chiral symmetry breaking. The decoupling of
massive fermions then destabilizes the Caswell-Banks-Zaks fixed point
$g^2_{\ast}$ in the gauge sector of the theory. The infrared of the theory is
then dominated by massless bosonic excitations, the Goldstone modes, and the
spectrum of the theory is characterized by a dynamically generated mass gap. A
similar reasoning also applies to QED${}_3$, see
e.~g.~\cite{Pisarski:1984dj,Appelquist:1986fd}.

These considerations suggest the existence of a {\it quantum critical point}
associated with a critical flavor number
$\Nf^{\text{CBZ}}\leq\Nfcr<\Nf^{\text{a.f.}}$ above which gauge theories
approach a conformally invariant limit in the infrared. Thus, $\Nf$ serves as
a control parameter for the quantum phase transition. 

Studies of the phase structure of strongly-flavored gauge theories have been performed employing
continuum methods as well as lattice simulations.  In QED${}_3$ many studies
have performed estimates to determine $\Nfcr$ using Dyson-Schwinger equations
and resummation
techniques~\cite{Pisarski:1984dj,Appelquist:1986fd,Appelquist:1988sr,Atkinson:1989fp,
  Pennington:1990bx,Curtis:1992gm,Burden:1990mg,Maris:1995ns,Gusynin:1995bb,Maris:1996zg,Fischer:2004nq}.
Since the dynamically generated mass is substantially smaller than the scale
set by the gauge coupling, lattice simulations of QED${}_3$ with many flavors
are remarkably
challenging~\cite{Dagotto:1989td,Hands:1989mv,Hands:2002dv,Hands:2004bh}.  The
phase structure of many-flavor QCD has also been studied employing continuum
methods~\cite{Caswell:1974gg,Banks:1981nn,Miransky:1996pd,Appelquist:1996dq,
  Appelquist:1997dc,Schafer:1996wv,Velkovsky:1997fe,Appelquist:1998rb,%
  Harada:2000kb,Sannino:1999qe,Harada:2003dc,Gies:2005as,Braun:2005uj,Braun:2006jd,Terao:2007jm,%
  Poppitz:2009uq,Armoni:2009jn,Sannino:2009qc,Sannino:2009me}, as well as lattice simulations
\cite{Kogut:1982fn,Gavai:1985wi,Fukugita:1987mb,Brown:1992fz,Damgaard:1997ut,%
  Iwasaki:2003de,Catterall:2007yx,Appelquist:2007hu,Deuzeman:2008sc,Deuzeman:2009mh,Appelquist:2009ty,%
  Fodor:2009wk,Fodor:2009ff,Pallante:2009hu,DeGrand:2010ba}. Recent results suggest in this
case that a conformal phase indeed exists with a quantum phase transition
occurring near
$9\lesssim\Nf^{\text{cr}}\lesssim 13$.

Given the existence of such a quantum critical point in an asymptotically free
gauge theory with $\Nf$ flavors, the question arises how the spectrum of the
theory behaves when we approach this quantum critical point from below. This
question is tightly bound to the question of the $\Nf$ dependence of the
dynamically generated scale associated with chiral symmetry breaking.  It is
well-known from studies of Dyson-Schwinger equations in the rainbow-ladder
approximation that physical observables, e.~g. the fermion condensate, exhibit
an exponential scaling close to $\Nfcr$, provided that the (momentum) scale
dependence of the gauge coupling can be
neglected~\cite{Miransky:1988gk,Appelquist:1996dq,Chivukula:1996kg,Appelquist:1998xf},
\be
\ksb\propto \Lambda\theta(\Nfcr-\Nf)\exp\left(
  {-\frac{\pi}{2\epsilon\sqrt{\alpha_1|\Nfcr
        - \Nf|} } 
}\right).\label{eq:1}
\ee
Here, $\ksb$ denotes a scale characteristic for the onset of symmetry breaking,
being directly proportional, say, to a symmetry-breaking condensate. The
quantities $\epsilon$ and $\alpha_1$ are pure constants arising from the
details of the theory and will be defined below. 
This behavior can be viewed as a
generalization of essential Berezinskii-Kosterlitz-Thouless (BKT)
scaling~\cite{Berezinskii,Berezinskii2,Kosterlitz:1973xp} to higher dimensional systems
\cite{Kaplan:2009kr}. We rush to add that the spectrum of the different
theories below and above $\Nfcr$ are substantially different. In particular, a
construction of an effective low-energy theory in terms of light scalar fields
may no longer be possible above $\Nfcr$.

Taking into account the running of the gauge coupling and going beyond the
standard rainbow-ladder approximation, the scaling behavior of physical
observables close to $\Nfcr$ has been analyzed
in~\cite{Braun:2005uj,Braun:2006jd,Braun:2009ns}. More precisely, the $\Nf$
dependence of a strict upper bound for the symmetry breaking scale has been
studied, scaling according to a power
law,
\be
\kcr \propto \Lambda |\Nfcr \!-\!\Nf|^{-\frac{1}{\Theta_0}}.\label{eq:2}
\ee
Here, $\kcr$ denotes the scale where the dynamics leading to symmetry breaking
becomes critical. This means that operators that trigger symmetry breaking
become relevant in an RG sense. As the system still has to run towards
  lower energy scales into the broken
phase, we have $\kcr> \ksb$, implying that \Eqref{eq:2} is an upper bound for
\Eqref{eq:1}. Near the critical flavor number the corresponding scaling
exponent is uniqely determined by the critical exponent $\Theta_0$ of the
gauge coupling at its infrared fixed point. This upper bound for the (chiral)
symmetry breaking scale can be translated into an upper bound for physical
observables~\cite{Braun:2009ns}.  In fact, the chiral-phase-transition
temperature as a function of the "external" control parameter $\Nf$ has been
computed with non-perturbative functional renormalization group methods. The
scaling of the {phase boundary has been found to be compatible with the analytically
derived scaling behavior~\cite{Braun:2005uj,Braun:2006jd}.}

Recently, the scaling behavior of physical observables has been investigated again
with the aid of Dyson-Schwinger equations in the rainbow-ladder approximation
also taking into account part of (momentum) scale dependence of the gauge coupling
by a proper adjustment of the scale~\cite{Jarvinen:2010ks}. It was then found that 
the exponential scaling behavior close to $\Nfcr$ of the symmetry breaking scale
is supplemented by a power-law behavior similar to the one found in 
Ref.~\cite{Braun:2005uj,Braun:2006jd,Braun:2009ns}.

In the present work, we aim to reveal the relation between these supposedly different
scaling laws and show rigorously what kind of scaling behavior we should expect 
close to the quantum critical point of asymptotically free gauge theories
with many flavors. Our arguments are based on very general RG considerations 
and involve only a few assumptions about the underlying fixed-point structure
of the theory under consideration. {In fact, we shall show that the above-given scaling laws
arise as two different limits of one and the same RG flow.} In addition, we point out the importance
of the scale-fixing procedure applied in the first place in order to compare
theories with different flavor numbers. As the scaling behavior of the low-energy
observables is accessible to a variety of nonperturbative methods, 
{we believe} that a rigorous understanding of scaling behavior near the phase
transition to the conformal phase will be very useful.

In Sect.~\ref{sec:miransky}, we briefly repeat the arguments that lead to 
an exponential scaling behavior at the quantum phase transition. 
In addition, we derive the leading-order correction
to the exponential scaling behavior. In Sect.~\ref{sec:powerlaw}, we then discuss the issue
of scale fixing underlying a meaningful comparison between theories with 
different flavor numbers. 
Moreover, we briefly review the arguments from Refs~\cite{Braun:2005uj,Braun:2006jd,Braun:2009ns}
which lead to a power-law-like scaling behavior for a strict upper bound for the (chiral)
symmetry breaking scale. In Sect.~\ref{sec:beyondmiransky} we then discuss the interrelation
of the scaling laws put forward in Refs.~\cite{Miransky:1988gk,Braun:2006jd,Braun:2009ns,Jarvinen:2010ks} and
derive the leading-order scaling behavior of a given infrared observable at the quantum critical point.
To illustrate our analytic findings, we present numerical results from a non-perturbative 
functional renormalization group study of the scaling behavior in many-flavor QCD in 
Sect.~\ref{sec:manyflavor}.

\section{Miransky scaling}\label{sec:miransky}
In this section we study exponential scaling behavior in gauge theories 
near a {\it quantum critical point}, also known as Miransky scaling~\cite{Miransky:1988gk,Miransky:1996pd}. 
\begin{figure*}[t]
\begin{center}
\begin{picture}(0,110)(80,20)
\put(-130,40){\includegraphics[scale=1]{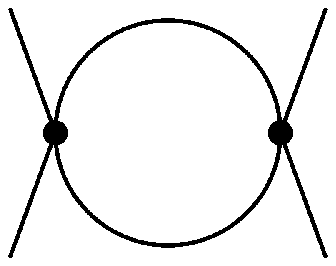}}
\put(-135,75){  \large$\lambda$}
\put(-45,75){  \large$\lambda$}
\put(-93,5){ \large$(a)$}
\put(30,40){\includegraphics[scale=1]{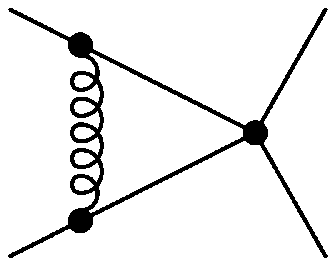}}
\put(45,40){  \large$g$}
\put(45,116){  \large$g$}
\put(108,75){  \large$\lambda$}
\put(67,5){ \large$(b)$}
\put(190,40){\includegraphics[scale=1]{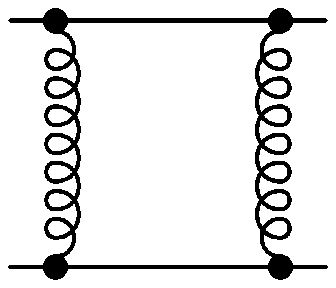}}
\put(197,30){  \large$g$}
\put(197,126){  \large$g$}
\put(263,126){  \large$g$}
\put(263,30){  \large$g$}
\put(227,5){ \large$(c)$}
\end{picture}
\end{center}
\caption{Representation of the terms on the right-hand side of the RG flow
  equation~\eqref{eq:4psiflow} by means of Feynman diagrams. Our
    functional RG studies, see Sect.~\ref{sec:manyflavor}, include
    resummations of all diagram types including ladder-diagrams generated by type (b) and (c)
    as well as the corresponding  crossed-ladder topologies.  }
\label{fig:feynman}
\end{figure*}

We shall keep our discussion as general as possible. For our purposes, however,
we consider a general class of theories where symmetry breaking and
  condensate formation is driven by fermionic self-interactions. Independently
  of whether these interactions may be fluctuation-induced (as in QCD) or
  fundamental, this class of theories can be parameterized by the following action:
\be
S_M &=& \int d^dx \Big\{ \bar{\psi}({\rm i}\partial\!\!\!\slash\; +
\bar{g}A\!\!\!\slash\;)\psi 
 + \bar{\lambda}_{\alpha\beta\gamma\delta} 
\bar{\psi}_{\alpha}\psi_{\beta}\bar{\psi}_{\gamma}\psi_{\delta}
\Big\}\,,
\label{eq:miransky_ansatz}
\ee
where $\alpha,\beta\,\dots$ denote a specific set of collective indices including, e.~g., flavor and/or color 
indices. In general, we expect to have more than just one four-fermion interaction channel, 
see e.~g. Sect.~\ref{sec:manyflavor} for QCD with many flavors. Note that symmetry
breaking is ultimately triggered by the interactions approaching criticality, i.e.,
becoming RG relevant. 
  
Here and in the following we do not allow for terms in the action which explicitly break the underlying
\mbox{symmetry}, such as current quark mass terms in QCD\footnote{The scaling behavior of observables 
with the (current) quark mass in the (quasi-)conformal phase of strongly-flavored gauge theories is of 
particular interest for lattice simulations and currently under investigation, 
see Refs.~\cite{DeGrand:2009mt,DelDebbio:2010ze,DelDebbio:2010jy}.}.

From the action~\eqref{eq:miransky_ansatz} we can derive the 
$\beta$ function of the dimensionless four-fermion coupling $\lambda$. It assumes the following simple form:
\be
\beta_{\lambda}\equiv\partial_t \lambda = (d-2)\lambda - a \lambda^2 -b\lambda g^2 -c g^4\,.
\label{eq:4psiflow}
\ee
Here, $t=\ln(k/\Lambda)$ denotes the RG 'time' with $k$ being the RG scale and
$\Lambda$ being a UV cutoff scale. The couplings
$\lambda\sim\bar{\lambda}/k^{(d-2)}$ and $g\sim\bar{g}/k^{4-d}$ denote
dimensionless and suitably renormalized couplings. The first term in
\Eqref{eq:4psiflow} arises from simple dimensional rescaling. The quantities
$a$, $b$ and $c$ do not depend on the RG scale but may depend on control
parameters, such as the number of quark flavors $\Nf$ or the number of colors
$\Nc$.\footnote{Note that the coefficients $a$, $b$ and $c$ can depend
  implicitly on the RG scale as soon as we introduce a dimensionful external
  parameter, e.~g., temperature $T$. However, the coefficients remain
  dimensionless since they depend only on the ratio $T/k$, see
  e.~g.~\cite{Braun:2005uj,Braun:2006jd}.} The various terms on the right-hand
side of Eq.~\eqref{eq:4psiflow} can be understood in terms of perturbative
Feynman diagrams~\cite{Braun:2006wu}, see Fig.~\ref{fig:feynman}. Equation
\eqref{eq:4psiflow} can also be derived from nonperturbative flow equations in
the limit of point-like (momentum-independent) interactions, {see Sect.~\ref{sec:manyflavor}.}  
Note that we have dropped terms proportional to the anomalous dimension of the
fermionic fields in Eq.~\eqref{eq:4psiflow}. We assume these contributions to
be small in the following. This is indeed the case in the chirally symmetric
regime of QCD, at least in the Landau gauge~\cite{Gies:2003dp}.

In Eq.~\eqref{eq:miransky_ansatz} we have not further specified the gauge
sector. In fact, let us ignore the running of the gauge coupling in this
section, and consider the gauge coupling as a {\it scale-independent}
"external" parameter. The RG flow of the gauge coupling is then trivially
governed by
\be
\partial_t g^2 \equiv 0\,.\label{eq:mcoup}
\ee
For example, this may be an acceptable approximation in the vicinity of an IR fixed 
point $g^2_{\ast}$. Note that the value of $g^2_{\ast}$ may depend on other control parameters
such as $\Nf$ or $\Nc$, cf. our discussion of
QCD with many flavors in Sect.~\ref{sec:manyflavor}. 

\begin{figure}[!t]
\begin{center}
\scalebox{0.5}[0.5]{
\begin{picture}(300,160)(80,40)
\includegraphics[scale=0.9]{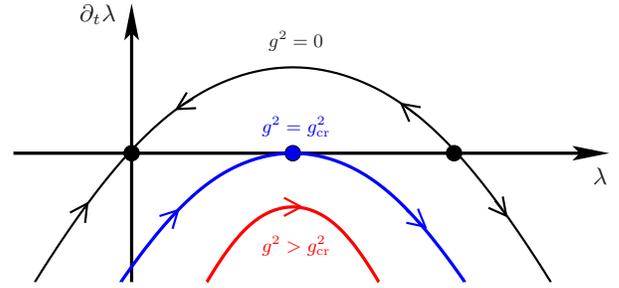} 
\Text(-10,+83)[c]{\scalebox{2}[2]{$\lambda$}}
\Text(-390,205)[c]{\scalebox{2}[2]{$\pat\lambda$}}
\Text(-240,185)[c]{\scalebox{1.6}[1.6]{$g^2=0$}}
\Text(-240,120)[c]{\scalebox{1.6}[1.6]{{\color{blue}$g^2 = g^2_{\rm cr}$}}} 
\Text(-240,30)[c]{\scalebox{1.6}[1.6]{\color{red}$g^2>g_{\text{cr}}^2$}}
\end{picture}
}
\end{center}
\medskip
\caption{Sketch of a typical $\beta$ function for the fermionic
  self-interactions $\lambda$, see \cite{Gies:2005as} and also \cite{Braun:2006jd} for
  a generalization to finite temperature): at zero gauge coupling, $g^2=0$ (upper
  black curve), the Gau\ss ian fixed point $\lambda=0$ is IR
  attractive. For $g^2 = g^2_{\rm cr}$ (middle/blue curve), the
  fixed-points merge due to a shift  of the parabola induced 
  by the gauge-field fluctuations $\sim g^4$. For gauge
  couplings larger than the critical coupling $g^2>g^2_{\text{cr}}$
  (lower/red curve), no fixed points remain and the
  self-interactions rapidly grow large, signaling chiral symmetry breaking.
  The arrows indicate the direction of the flow towards the infrared.
} 
\label{fig:parab}
\end{figure}

In Fig.~\ref{fig:parab} we show a sketch for the $\beta_{\lambda}$ function,
implicitly assuming that $a>0$, $b>0$ and $c>0$ in Eq.~\eqref{eq:4psiflow}. For a
vanishing gauge coupling $g^2$ we find two fixed points, an IR attractive
Gaussian fixed point at $\lambda=0$ and an IR repulsive fixed point at
$\lambda>0$. For increasing $g^2$ these fixed points approach each other and
eventually merge for a {\it critical} value $g^2_{\rm cr}$, 
\be
g_{\rm cr}^2=\frac{d-2}{b + 2 \sqrt{ac}}\,.  
\ee
  For $g^2 >g^2_{\rm cr}$ the four-fermion coupling then becomes a relevant operator and
increases rapidly towards the IR indicating the onset of (chiral) symmetry
breaking. Thus, the four-fermion coupling $\lambda$ necessarily\footnote{Here,
  we assume that the initial conditions at the UV scale $k=\Lambda$ for the
  four-fermion coupling $\lambda$ are chosen such that $\lambda_{\Lambda}$ is
  smaller than the value of the IR repulsive fixed point, see
  Fig.~\ref{fig:parab}.  In beyond-standard model applications
  $\lambda_{\Lambda}$ is sometimes considered to be a finite parameter, see
  e.~g.~\cite{Fukano:2010yv}. We therefore add that the exponential scaling
  behavior discussed below can only be observed when $\lambda_{\Lambda}$ is
  chosen to be smaller than the value of the repulsive fixed point for a given
  $g^2$.  Otherwise, we expect a power-law-like scaling
  behavior~\cite{Braun:2010tt}.  } diverges for $g^2 > g^2_{\rm cr}$ at a
finite RG scale $\ksb=\ksb(g^2)$. This divergence is, of course, an artifact
of the over-simplistic approximation \eqref{eq:miransky_ansatz}, but can be
related to a symmetry-breaking transition in the {effective 
Landau-Ginzburg-type} potential for
fermion-bound states. {Even though $k_{\text{SB}}$ is not a direct observable}, it
sets the scale for observables such as condensates, decay constants,
critical temperatures, etc.  {This picture of the emergence of chiral symmetry} has been put
forward in~\cite{Gies:2005as,Braun:2005uj,Braun:2006jd,Braun:2009ns} and
successfully employed for an anlysis of the phase structure of QCD with
various numbers of flavors and colors at zero and finite
temperature~\cite{Gies:2005as,Braun:2005uj,Braun:2006jd,Braun:2009ns}.
Moreover, this picture has also been employed to study conformal scaling in
quantum field theories, see e.~g. Ref.~\cite{Kaplan:2009kr}.

Let us now briefly discuss the scaling behavior of the symmetry-breaking scale $\ksb$ when
$g^2$ is varied by hand as a constant "external" parameter. To this end, we have to solve the RG flow 
equation~\eqref{eq:4psiflow}. We find:
\be
\ln k -\ln\Lambda=-\frac{2\arctan\left(\frac{bg^2 -(d-2) + 2a\lambda^{\prime}}
{\delta(g^2)}\right)}{\delta(g^2)} 
\,\Bigg|^{\lambda}_{\lambda_{\rm UV}}\,.
\label{eq:mlambda_sol}
\ee
with 
\be
\delta (g^2)=\sqrt{4acg^4 - ((d-2) - bg^2)^2}\,.
\ee
Here and in the following we assume $b>0$ without loss of generality.
From Eq.~\eqref{eq:mlambda_sol}, we obtain $\ksb$ by solving for the zero 
of $1/\lambda(k)$, i. e. $1/\lambda(\ksb)=0$:
\be
\ln\ksb -\ln\Lambda = -\frac{\pi}{\delta(g^2)} + {\rm const.}\,.\label{eq:m_general}
\ee
Here, we have chosen the initial conditions such that $\lambda_{\rm UV}=\lambda_{\rm max}$ where
$\lambda_{\rm max}$ denotes the position of the maximum of the
$\beta_{\lambda}$ function, i.e., the peak of the parabola in
  Fig.~\ref{fig:parab}. An expansion of~\eqref{eq:m_general} around $g^2_{\rm cr}$ yields\footnote{Note that $g^2_{\rm cr}$
is defined to be the value of $g^2$ for which the $\beta_{\lambda}$ function has exactly one zero. In
general there exist two solutions for $g^2_{\rm cr}$, however, one of which can be excluded from a physical
point of view.}
\be
\ksb \propto \Lambda \theta(g^2-g^2_{\rm cr}) 
\exp\left( {-\frac{\pi}{2\epsilon\sqrt{(g^2 - g^2_{\rm cr})}  }}
\right)\,,
\label{eq:miransky}
\ee
where $\epsilon$ is a numerical factor,
\be
\epsilon = \sqrt{\frac{(d-2)(2ac+b\sqrt{ac})}{b+2\sqrt{ac}}}\,,
\label{eq:defeps}
\ee
which in general depends on the details of the theory under consideration, 
e.~g. the number of colors and flavors in QCD,
In any case, we find an exponential Miransky-scaling behavior of $\ksb$ for
$g^2$ close to $g^2_{\rm cr}$.  Since the dynamically generated scale $\ksb$
sets the scale for the low-energy sector, we expect that all IR observables
$\mathcal O$ scale according to:
\be
{\mathcal O}=f_{\mathcal O}\,\ksb^{d_{\mathcal O}}\,,\label{eq:ObsScaling}
\ee
where $d_{\mathcal O}$ is the canonical mass dimension of the observable
$\mathcal O$ and $f_{\mathcal O}$ is a function which does not depend on
$g^2_{\rm cr}$ but may depend on $g^2$ and other external parameters, e.~g.,
$\Nf$ and/or $\Nc$. The function $f_{\mathcal O}$ can be computed
systematically within certain approximations schemes such as large-$\Nc$
expansions or chiral perturbation theory, see
e.~g.~\cite{Braun:2009ns,Jarvinen:2010ks}.
In the context of QCD the scaling law in Eq.~\eqref{eq:miransky} has been
first derived by Miransky~\cite{Miransky:1988gk,Miransky:1996pd} but has also
been found in the context of specific 2-dimensional condensed-matter
systems~\cite{Kosterlitz:1973xp}. The derivation of the scaling
law~\eqref{eq:miransky} via an analysis of the RG flow of four-fermion operators has been recently
pointed out by Kaplan, Lee, Son and Stephanov~\cite{Kaplan:2009kr}.

Let us now briefly discuss the consequences of the
scaling law~\eqref{eq:miransky} when we apply our considerations 
to strongly-flavored gauge theories, such as QCD with many quark
flavors or ${\rm QED}_3$. In these cases we may choose the IR fixed-point of the gauge coupling as an
external parameter, i.~e. $g^2=g^2_{\ast}(N_f)$ in Eq.~\eqref{eq:miransky}. Depending on the $\Nf$ 
dependence of the coefficients $a$, $b$ and $c$ in the $\beta_{\lambda}$ function, the critical
value for the gauge coupling may depend on the number of flavors as well, $g^2_{\rm cr}=g^2_{\rm cr}(\Nf)$.
The critical number of quark flavors $\Nfcr$ can then be obtained from the
criticality condition
\begin{equation}
g^2_{\rm cr}(\Nfcr)=g^2_{\ast}(\Nfcr).\label{eq:critcond}
\end{equation}
This corresponds to the coupling value for which the two fixed points
of the four-fermion coupling $\lambda$ merge and then annihilate each other for $g^2> g_{\text{cr}}^2  $.
Expanding $g^2_{\ast}(\Nf)-g^2_{\rm cr}(\Nfcr)$ around $\Nfcr$,
\begin{equation}
 g^2_{\ast}(\Nf)-g^2_{\rm cr}(\Nfcr)
= \alpha_1(\Nf\!-\!\Nfcr)+\alpha_2(\Nf\!-\!\Nfcr)^2+\dots,\label{eq:g2g2cr}
\end{equation}
and plugging~\eqref{eq:g2g2cr} into~\eqref{eq:miransky}, we find the
exponential $\Nf$ scaling of $\ksb$:
\begin{equation}
\ksb\propto \Lambda\theta(\Nfcr-\Nf)\exp\left(
  {-\frac{\pi(1-\frac{\alpha_2}{\alpha_1}|\Nfcr\!-\!\Nf|+\dots)}{2\epsilon\sqrt{\alpha_1|\Nfcr\!
        - \!\Nf|} } 
}\right).\label{eq:mcorrections}
\end{equation}
Whether the size of the regime for exponential scaling is small depends on the
ratio $\alpha_2/\alpha_1$ which in turn depends on the theory under consideration. Thus, the size
of the scaling regime may presumably be different in, e.~g., QCD and ${\rm QED}_3$. In Sect.~\ref{sec:manyflavor}
we compare the analytic findings of this section with results from a numerical analysis of 
QCD with many flavors.

\section{Power-law scaling}\label{sec:powerlaw}
\begin{figure*}[t]
\begin{center}
\begin{picture}(0,130)(80,40)
\put(-130,40){\includegraphics[scale=0.336]{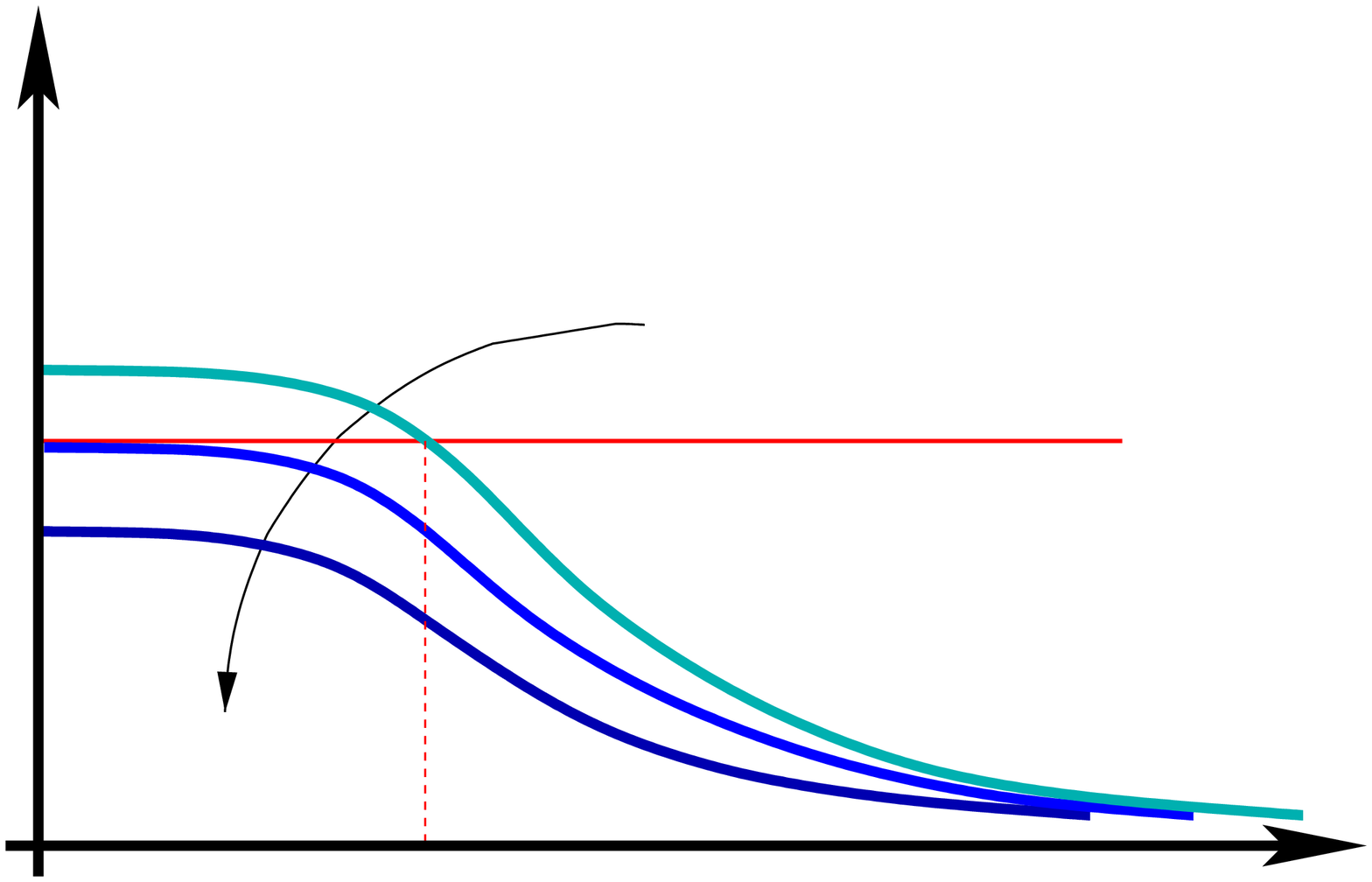}}
\put(-140,170){\large $g^2$}
\put(-150,100){ \color{red} \large$g^2_{\rm cr}$}
\put(70,30){ \large$k$}
\put(-75,30){\color{red} \large$k_{\rm cr}$}
\put(-55,130){ \large$\Nf$}
\put(95,40){\includegraphics[scale=0.55]{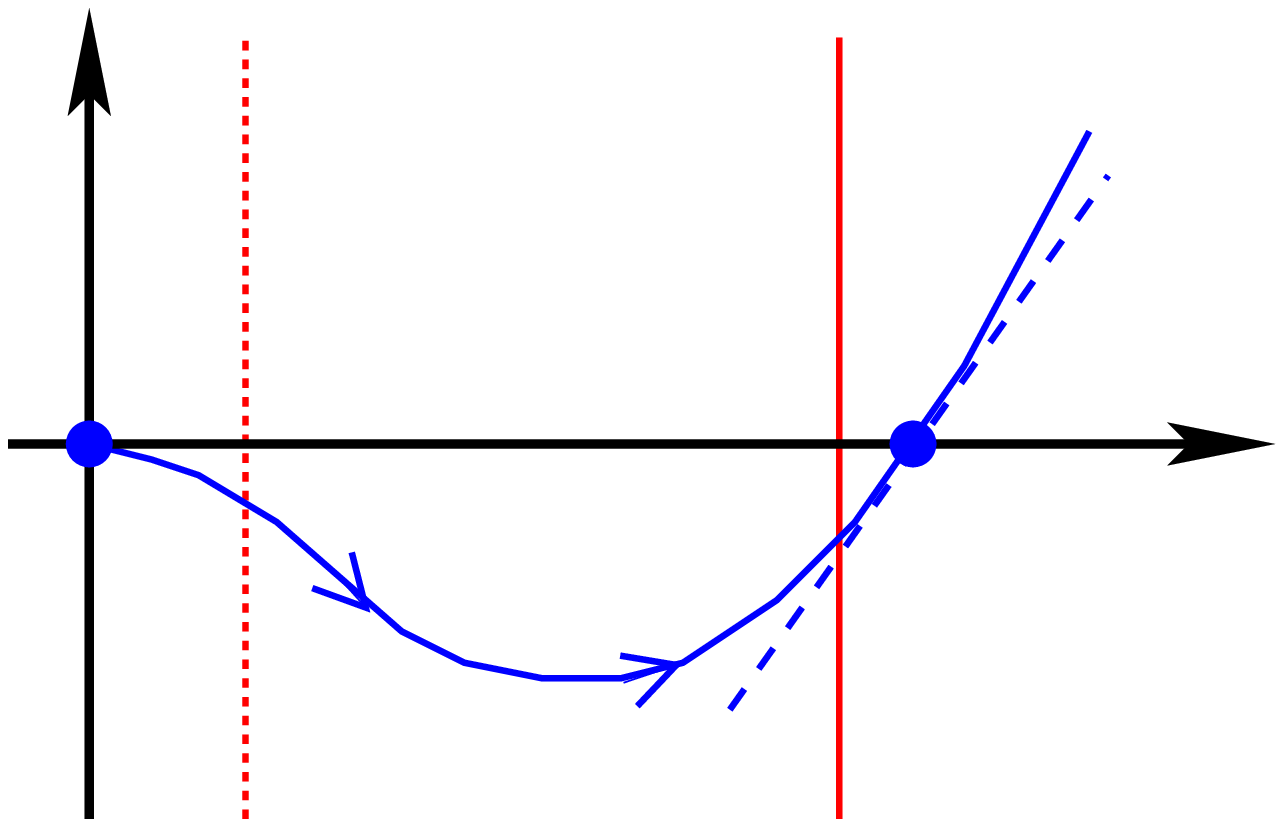}}
\put(80,170){\large $\partial_t g^2$}
\put(295,85){ \large$g^2$}
\put(130,25){\color{red} \large$g^2_{\Lambda}$}
\put(225,25){\color{red} \large$g^2_{\rm cr}$}
\put(245,85){\color{blue} \large$g^2_{\ast}$}
\put(275,130){\color{blue} \large$|\Theta|$}
\end{picture}
\end{center}
\caption{Left panel: illustration of the IR running of the gauge coupling in
  comparison with the critical value of the gauge coupling $g_{\text{cr}}^2$. Below the
  conformal window, $\Nf<\Nfcr$, $g^2$ exceeds the critical value $g^2_{\rm cr}$, 
  triggering the approach to \xsb. For increasing flavor number, the IR fixed-point 
  value $g_{\ast}^2$ becomes smaller than the critical value, indicating that the theory
  is inside the conformal window. Right panel: sketch of the $\beta$ function of the running
  gauge coupling. The slope of the $\beta$ function at the IR fixed-point
  corresponds to minus the critical exponent $\Theta$, cf. Eq.~(\ref{eq:FPR}). The vertical
  line to the right gives the value of $g^2_{\rm cr}$. {The dotted line} gives 
  the value of the gauge coupling at the UV scale $\Lambda$ which we keep fixed for
  all $\Nf$. By contrast, the value of $g^2_{\rm cr}$ may depend on $\Nf$.
  The arrows indicate the direction of the flow towards the infrared.
}
\label{fig:sketches}
\end{figure*}

In this section we discuss how the running of the gauge coupling affects the
RG flow of the four-fermion coupling(s). In particular, we argue that (chiral)
symmetry breaking in strongly-flavored gauge theories is a multi-scale
problem, in contrast to the scenario discussed in the previous section.  In
other words, the (chiral) symmetry breaking scale $\ksb$ discussed above and
its scaling with the control parameters, e. g. the number of flavors $\Nf$,
depends on the scale fixing and its potential flavor dependence.

In the following, we include the running of the gauge coupling which goes beyond 
standard rainbow-ladder approaches employed in the context of strongly-flavored gauge
theories, see e.~g. Ref.~\cite{Appelquist:1998rb}. 

Since we are eventually interested in the scaling behavior of IR observables, e. g.
the fermion condensate, it is important to realize that a variation of the 
flavor number does not quite correspond to a change of a parameter of the 
theory. It rather corresponds to changing the theory itself. We would like to
stress that there is indeed no unique way to unambiguously compare theories 
of different flavor number with each other, as different theories may be fixed
at different scales.

As we have argued in more detail in Ref.~\cite{Braun:2009ns}, fixing the scale
of theories with, say, different flavor numbers $\Nf$ by keeping the running
coupling at some scale $\Lambda$ (e.~g. $\tau$ mass) fixed to a certain
value, seems to be a well accessible prescription for many non-perturbative
methods. In general, it is important to take care that this scale-fixing
procedure is not (or as little as possible) spoilt by scheme dependencies. The
latter constraint essentially rules out $\LQCD$ as a proper scale in QCD to be
kept fixed in theories with different flavor numbers.  Of course, it is also
possible, e.~g. in lattice QCD simulations, to keep the value of an IR
observable fixed for theories with different $\Nf$, e.~g. the pion decay
  constant or the critical temperature. We shall briefly comment on such a procedure below. For what
follows, however, we choose a mid-momentum scale for the scale fixing, lying
in between the high-scale perturbative running and the more interesting
non-perturbative dynamics.  Thus, we fix the theories at any $\Nf$ by keeping
the running coupling at some intermediate scale $\Lambda$ fixed to a
certain value, say $\alpha_{\Lambda}$.

To be more specific, we shall focus our discussion on strongly-flavored
asymptotically free gauge theories, such as QCD with many flavors and ${\rm
  QED}_3$.\footnote{By asymptotic freedom, we refer to the vanishing of the
  dimensionless renormalized coupling in the UV.} In such theories, the
dependence of the running coupling on the scale and on further control
parameters such as $\Nf$ is expected to modify Miransky scaling. In
particular, an understanding of the universal scaling behavior of observables
in the ordered phase close to the phase transition at $\Nfcr$ is of interest.
However, the arguments also apply to other
theories in which dynamical chiral symmetry breaking is trigged by a running
coupling which approaches a non-trivial IR fixed point.

For a monotonically increasing coupling flow, the value of the non-trivial IR
fixed point $g_\ast^2$ of the gauge coupling corresponds to the maximum
possible coupling strength of the system in the conformal window, i.~e. for
$\Nfcr<\Nf<\Nf^{\rm a.f.}$.  As both $g_\ast^2$ and $g_{\text{cr}}^2$ depend
on the number of flavors, the criticality condition
$g_\ast^2(\Nfcr)=g_{\text{cr}}^2(\Nfcr)$ defines the lower
end of the conformal window and thus the critical flavor number, see
Sect.~\ref{sec:miransky} and the left panel of Fig.~\ref{fig:sketches} for an
illustration.

For $g_\ast^2>g_{\text{cr}}^2$, our model~\eqref{eq:miransky_ansatz} is below the conformal window and
runs into the broken phase. Slightly below the conformal window, the running
coupling $g^2$ exceeds the critical value while it is in the attractive domain
of the IR fixed point $g_\ast^2$. The flow in this fixed-point regime can approximately 
be described by a $\beta$ function expanded around the fixed point $g_\ast^2$:
\begin{equation}
\beta_{g^2}\equiv \pat g^2 =-\Theta\, (g^2\!-\! g_\ast^2)\!+\!
{\mathcal O}((g^2\! -\! g_\ast^2)^2)\,. \label{eq:FPR}
\end{equation}
The universal "critical exponent'" $\Theta$ denotes (minus) 
the first expansion coefficient. We know that $\Theta<0$, since the fixed point is IR
attractive, see right panel of Fig.~\ref{fig:sketches}. In general, the critical exponent depends 
on $\Nf$, $\Theta=\Theta(\Nf)$.
The solution to Eq.~\eqref{eq:FPR} for the running coupling in 
the fixed-point regime simply reads
\begin{equation}
g^2(k)=g_\ast^2-\left(\frac{k}{k_0}\right)^{-\Theta}, \label{eq:FPsol}
\end{equation}
where the scale $k_0$ is implicitly defined by a suitable initial condition
and is kept fixed in the following as we keep the UV scale ${\Lambda}$ fixed.

The scale $k_0$ corresponds to a scale where the system
is already in the fixed-point regime. 
For the present fixed-point considerations, $k_0$ 
provides for all dimensionful scales. However, from the knowledge of the full RG
trajectory, $k_0$ can be related to the initial scale $\Lambda$, say the $\tau$ mass
scale in QCD, by RG evolution.

A necessary condition for (chiral) symmetry breaking is that $g^2_\ast>
g^2_{\text{cr}}$.  This implies that $g^2(k)$ 
exceeds $g_{\text{cr}}^2$ at some scale $k_{\text{cr}}$ which is implicitly
defined by the criticality condition,
$g_\ast^2(\Nfcr)=g_{\text{cr}}^2(\Nfcr)$, and therefore 
\be
\kcr \geq \ksb\,,
\ee
where $\ksb$ is the scale at which the four-fermion coupling $\lambda$ diverges, see 
Sect.~\ref{sec:miransky}. Thus, $\kcr$ is an upper bound for the symmetry breaking
scale $\ksb$. From Eq.~\eqref{eq:FPsol} and the criticality condition
$g^2(k_{\text{cr}})= g^2_{\text{cr}}$, we derive an estimate for $\kcr$ valid in the
fixed-point regime
\begin{equation}
k_{\text{cr}}\simeq k_0\, (g_\ast^2
-g_{\text{cr}}^2)^{-\frac{1}{\Theta}}. \label{eq:kcrest}
\end{equation}
The scale $\kcr$ is dynamically generated. Note that
$k_{\text{cr}}/k_0\to 0$ for $g^2_\ast\to g^2_{\text{cr}}$ from above. Due to
our scale-fixing procedure, this scale depends on $\Nf$ and $\Nfcr$ in a
non-trivial way\footnote{Note that it is, in principle, possible to adjust the
  initial value of the coupling at the initial scale such that the scale
  $\kcr$ is independent of $\Nf$ and $\Nfcr$. As indicated above, {we expect that} 
  such a scale-fixing procedure would, however, be strongly affected by
  scheme-dependencies at least in our truncation.}. Using
Eq.~\eqref{eq:g2g2cr} and a Taylor expansion of the critical exponent near the
quantum phase transition,
\be
\Theta(\Nf)=\Theta_0 + \Theta_1 (\Nf-\Nfcr) +{\mathcal O}((\Nf-\Nfcr)^2)\,,
\label{eq:taylor}
\ee
we find the following $\Nf$ dependence of $\kcr$ for $\Nf\leq \Nfcr$:
\begin{eqnarray}
\kcr &\simeq& k_0|\Nfcr \!-\!\Nf|^{-\frac{1}{\Theta_0}} \label{eq:kcr}\\
&&\times\left( 1 
\!-\! \frac{ |\Nfcr\! -\!
\Nf|}{\Theta_0}\left( \frac{\alpha_2}{\alpha_1} 
-\frac{\Theta_1}{\Theta_0}\ln (\alpha_1 |\Nfcr\! -\! \Nf|)\right)\right) 
\nonumber\\
&&+\dots\,,\nonumber
\end{eqnarray} 
where $\Theta_0=\Theta(\Nfcr)$. Since $\kcr$ defines the scale at which the fixed-points in
the $\beta$ function of the four-fermion coupling merge, the existence of a
finite $\kcr$ can be considered as a necessary condition for (chiral) symmetry
breaking. Thus, we expect that the scale for a given IR observables ${\mathcal
  O}$ for $\Nf\leq\Nfcr$ is set by $\kcr$:
\be
{\mathcal O}= f_{\mathcal O} \kcr^{d_{\mathcal O}}\,,
\ee
where $d_{\mathcal O}$ is the canonical mass dimension and $f_{\mathcal O}$
dependes on $\Nf$ but not on $\Nfcr$, see also
Eq.~\eqref{eq:ObsScaling}. However, we would like to stress that $\kcr$ does
not include the full dependence of $\ksb$ on $(\Nf-\Nfcr)$, i. e.  $\kcr/\ksb
\neq {\rm const.}$ is still a function of the control parameter, as we shall
discuss in the subsequent section.

In Ref.~\cite{Braun:2005uj,Braun:2006jd,Braun:2009ns} we have implicitly used the existence of 
a finite $\kcr$ to estimate
the chiral phase transition temperature in QCD as a function of $\Nf$. 
For a given value of $\Nf$ the phase transition temperature 
is given by the highest temperature for which we still have $\kcr>0$. 
We have indeed found that $T_{\rm cr}$ scales according to Eq.~\eqref{eq:kcr}:
\be
T_{\text{cr}}\sim k_0 |\Nfcr -\Nf|^{-\frac{1}{\Theta_0}}.
\label{eq:TcrTheta}
\ee
Strictly speaking, this is only an upper bound for the chiral phase transition temperature since
it is only sensitive to the emergence of a fermion condensate on intermediate (momentum)
scales but insensitive to a fate of the condensate in the deep IR close to $T_{\rm cr}$ due to 
fluctuations of Goldstone modes~\cite{Braun:2009si}, also known as a local ordering phenomena.
Such strong IR fluctuations of the Goldstone modes may yield corrections to the scaling law for
the critical temperature given above\footnote{We would naively expect that corrections to Eq.~\eqref{eq:TcrTheta} 
can be only resolved in lattice simulations with very small masses for the 
pseudo Goldstone modes and on very large lattice sizes, see also Sect.~\ref{sec:beyondmiransky}.}. 
Nevertheless, relation \eqref{eq:TcrTheta} is an analytic prediction for 
the shape (of the upper bound of) the chiral phase boundary in the ($T,\Nf$) 
plane.

At vanishing temperature, the analysis of the scaling behavior of IR
observables is simplified compared to a scaling analysis at finite temperature
since dimensional reduction does not set in in the deep IR {enhancing
the Goldstone modes.} {Based on the observed scaling behavior of $\kcr$ with the
number of flavors, we are therefore in a position} to derive the $\Nf$
scaling of low-energy observables, such as fermion condensates, at zero
temperature.

\section{Beyond Miransky scaling}\label{sec:beyondmiransky}
Let us now discuss how the symmetry breaking scale $\ksb\leq \kcr$ depends on
$(\Nf-\Nfcr)$. We consider again a Lagrangian of the
form~\eqref{eq:miransky_ansatz}, and assume that $\Nf\lesssim\Nfcr$. The
crucial new ingredient compared to the derivation of Miransky scaling is the
RG flow of the coupling.  We also assume that the system has already evolved
from the initial UV scale ${\Lambda}$ to the scale $\kcr$ at which the fixed
points of the $\beta$ function of the four-fermion coupling have
merged. Sufficiently close to $\Nfcr$, the flow of the gauge coupling is
governed by the fixed point regime for $g^2 > g^2_{\rm cr}$.  The running of
the gauge coupling is then given by (cf. \Eqref{eq:FPsol})
\begin{eqnarray}
g^2(k) &=& g^2_{\ast} - (g^2_{\ast}-g^2_{\rm
  cr})\left(\frac{k}{\kcr}\right)^{-\Theta} \nonumber\\
&= &g^2_{\ast} - (\Delta g^2)\left(\frac{k}{\kcr}\right)^{-\Theta}\,,\label{eq:g2fp}
\end{eqnarray}
where $\Delta g^2=g^2_\ast-g^2_{\text{cr}}$.  Recall that $g^2_{\ast}\sim \Nf$
and $\Delta g^2\sim |\Nfcr-\Nf|$. Plugging Eq.~\eqref{eq:g2fp} into
Eq.~\eqref{eq:4psiflow}, we find
\be
\beta_{\lambda}&\equiv&\partial_t\lambda=\beta_{\lambda}\Big|_{g^2_{\ast}}
+\frac{\partial \beta_{\lambda}}{\partial g^2}\Big|_{g^2_{\ast}}(\Delta g^2)\left(\frac{k}{\kcr}\right)^{-\Theta}
+\dots\label{eq:betacorr}\\
&=&\! (d\!-\!2)\lambda - a \lambda^2 -b\lambda g^2_{\ast} -c g^4_{\ast}
+\frac{\partial \beta_{\lambda}}{\partial g^2}\Big|_{g^2_{\ast}}\left(\!\frac{k}{k_0}\!\right)^{-\Theta}
\!\!\!\!\!\! +\dots, \nonumber
\ee
where we have used Eq.~\eqref{eq:kcrest}. Recall that $k\leq \kcr \ll k_0$ and
$\Theta < 0$.  We observe that the zeroth order in $\Delta g^2$ coincides with
the $\beta_{\lambda}$ function for which we have found an (implicit) analytic
solution for constant $g^2$ in Sect.~\ref{sec:miransky} yielding Miransky
scaling. We refer to this analytic solution as $\lambda_{g^2_{\ast}}$. The
solution of the $\beta$~function~\eqref{eq:betacorr} can then be found by an
expansion around the solution $\lambda_{g^2_{\ast}}$:
\be
\lambda &=& \lambda_{g^2_{\ast}} + (\Delta g^2)\delta\lambda + \dots \nn\\
&=&\lambda_{g^2_{\ast}} + \left(\frac{\kcr}{k_0}\right)^{-\frac{1}{\Theta}}\delta\lambda + \dots
\,.
\ee
This allows us to systematically compute the scaling behavior for $\Nf\lesssim\Nfcr$. Since we are interested
in the (chiral) symmetry breaking scale $\ksb$ we have to solve $1/\lambda(\ksb)=0$ for $\ksb$. In zeroth order
the scale $\ksb$ can be computed along the lines of our analysis in Sect.~\ref{sec:miransky}. We find
\be
\ksb &\propto& \kcr \theta(\Nfcr-\Nf) \exp\left( {-\frac{\pi}{2\epsilon\sqrt{\alpha_1|\Nfcr\! -\! \Nf|} }
}\right)\nn\\
&\simeq& k_0 \theta(\Nfcr-\Nf)|\Nfcr -\Nf|^{-\frac{1}{\Theta_0}}\nn\\
&&\times \exp\left( {-\frac{\pi}{2\epsilon\sqrt{\alpha_1|\Nfcr\! -\! \Nf|} }}\right)
,
\label{eq:ksbscaling}
\ee 
where we have used Eq.~\eqref{eq:kcr} in leading order. 
Higher order corrections to Eq.~\eqref{eq:ksbscaling} can be computed systematically
as outlined above and in the previous sections. Thus, we have found a {\it universal} correction to the exponential scaling 
behavior which is uniquely determined by the universal "critical" exponent $\Theta$. A similar result 
has been suggested very recently by Jarvinen and Sannino using a standard rainbow-ladder approach with a 
constant gauge coupling but a properly adjusted scale~\cite{Jarvinen:2010ks}. Our RG analysis demonstrates
in a simple and systematic way that such a rainbow-ladder approach is indeed justified and yields the correct 
leading-order scaling behavior.

Let us now turn to the scaling behavior of physical observables.  The scale of
all low-energy observables is set by $\ksb$. In other words, $\ksb$ represents
the UV cutoff of an effective theory at low energies, such as chiral
perturbation theory, quark-meson or NJL-type models in case of QCD. At zero
temperature we therefore expect that a given IR observable $\mathcal O$ with
mass dimension $d_{\mathcal O}$ scales according to
\be
{\mathcal O}=f_{\mathcal O}(\Nf)\theta(\Nfcr-\Nf)\,\ksb^{d_{\mathcal O}}\,,\label{eq:slawcorr}
\ee
where $f_{\mathcal O}(\Nf)$ is a function which depends on $\Nf$ but not on $\Nfcr$. As mentioned above, 
$f_{\mathcal O}(\Nf)$ can be in principle systematically computed in QCD using, e.~g., chiral perturbation
theory or a large-$\Nc$ expansion. For instance, in a large-$\Nc$ expansion it is straightforward 
to derive the leading $\Nf$ dependence of the 
function $f_{\mathcal O}(\Nf)$ for the pion decay constant $f_{\pi}$. It reads~\cite{Braun:2009ns}
\be
f_{f_{\pi}}(\Nf)\sim\sqrt{\Nf}\,.
\ee

The scaling law~\eqref{eq:slawcorr} together with (\ref{eq:ksbscaling}) 
represents one of the main results of this
work. It can be used as an ansatz to fit, e.~g., data from lattice
simulations.  This scaling law is remarkable for a number of reasons: first,
it relates two universal quantities with each other: quantitative values of
observables and the IR critical exponent. Second, it establishes a
quantitative connection between the (chiral) phase structure and the IR gauge
dynamics ($\Theta$). Third, it is a parameter-free prediction following
essentially from scaling arguments. {Moreover, it shows that Miransky scaling
and power-law scaling are simply two limits of the very same set of RG flows:
in the limit $\Theta\to\infty$ we find pure Miransky-scaling behavior, while we have
pure power-law scaling in the limit $\Theta\to 0$.}

{At this point, we would like to emphasize once more that the scaling
  behavior of any IR observable near $\Nfcr$ depends crucially on the
  scale-fixing procedure applied in the first place. Still, the
  universal scaling will always show up at one or the other place and thus
  cannot be removed, as stressed in Ref.~\cite{Braun:2009ns}. Our choice to
  fix the scale at $m_\tau$ which is large enough not to be affected by 
  chiral-symmetry-breaking is certainly not unique. In principle, the point
  where to fix the scale can be chosen as a free function of $\Nf$. 
  In \Eqref{eq:FPsol}, this would correspond to the choice of an arbitrary 
  function $k_0=k_0(\Nf)$ for the global scale, which then appears also in 
  the scaling relations~\eqref{eq:kcr}, \eqref{eq:TcrTheta} and 
  \eqref{eq:ksbscaling}. Indeed, an extreme choice would be given by
  measuring all dimensionful scales in units of a scale induced by chiral
  symmetry breaking (such as $T_{\rm cr}$ or $f_{\pi}$). In this case, all
  chiral observables would jump non-analytically across
  $\Nf=\Nf^{\text{cr}}$. Nevertheless, the scaling relations would then
  translate into scaling relations for other non-chiral external scales: e.g.,
  the scale $k$ at which the running coupling acquires a specific value (say
  $\alpha=0.322$) would diverge with $\Nf\to\Nf^{\text{cr}}$ according to
  $k\sim |\Nf -\Nf^{\text{cr}}|^{-\frac{1}{|\Theta_0|}}$. This point of view
  can constitute a different way of verifying our scaling relations on the
  lattice.}

Let us conclude this section with a discussion of the importance of the corrections to the exponential scaling
behavior due to the running of the gauge coupling. 
To that end, it is convenient to consider the logarithm of the (chiral) symmetry breaking scale $\ksb$,
\begin{equation}
\ln \ksb = {\rm const.} - \frac{1}{\Theta_0}\ln |\Nfcr -\Nf| -\frac{\pi}{2\epsilon\sqrt{\alpha_1|\Nfcr\! -\! \Nf|}}\,.
\end{equation}
This expression  can be used to estimate the regime in which the corrections to the exponential scaling become subdominant.
For this, we compute the minimum of the function
\be
\frac{1}{|\Theta_0|}\ln |\Nfcr -\Nf| + \frac{\pi}{2\epsilon\sqrt{\alpha_1|\Nfcr\! -\! \Nf|}}
\ee
with respect to $|\Nfcr -\Nf|$. In accordance with \Eqref{eq:taylor}, we
  assume $|\Nfcr -\Nf|<1$ here. From this, we can then estimate that corrections to the 
exponential scaling behavior are subdominant as long as
\be
|\Nf -\Nfcr|\lesssim \frac{\pi^2 |\Theta_0|^2}{16 \epsilon^2 \alpha_1}\,,\label{eq:size}
\ee
with $\epsilon$ begin defined in \Eqref{eq:defeps}.  We observe that
corrections to Miransky scaling due to the running of the gauge coupling are
small when $|\Theta_0|\gg 1$ and large when $|\Theta_0|\ll 1$.  To be more
specific, let us consider QCD with many flavors: assuming $\Nfcr\approx 12$,
we extract $\Theta_0\approx 0.4$ from the two-loop $\beta_{g^2}$
function. {From Eq.~\eqref{eq:size} the region where pure Miransky scaling dominates is 
then found to be confined to the regime $|\Nf-\Nfcr|\lesssim 0.3$.} Thus, we
expect that the exponential scaling behavior is dominantly visible only very close to
$\Nfcr$. According to this estimate, the $\Theta$-dependent {universal corrections are therefore
more significant in QCD.}

\begin{figure}[!t]
\includegraphics[scale=0.6]{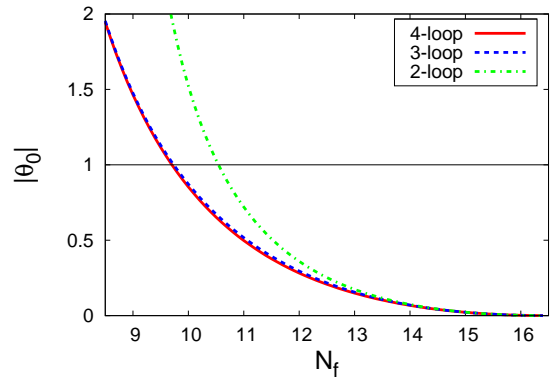} 
\caption{Critical exponent $\Theta$ of the running gauge coupling at the Caswell-Banks-Zaks fixed point
as a function of the number of flavors $\Nf$ as obtained from two-, three- and four-loop perturbation theory
in the $\MSbar$ scheme.
} 
\label{fig:Theta}
\end{figure}
In QCD, it appears to be a general feature  that $\Theta_0$ decreases with
$\Nfcr$. Estimates of $\Theta_0$ within two- and higher-loop approximations in the $\MSbar$ 
scheme are summarized in Fig.~\ref{fig:Theta}. Therefore, power-law
scaling is more prominent for larger $\Nfcr$. In particular, power-law scaling
should be visible if theories are probed only for integer values of $\Nf$ as,
e.g., on the lattice. 

The role of $|\Theta|$ for the scaling behavior close to $\Nfcr$ can also be
understood by simply looking at the $\beta _{g^2}$ function of the gauge
coupling, see Fig.~\ref{fig:sketches}. For $|\Theta|\gg 1$ the gauge coupling
runs very fast into its IR fixed point once it has passed $g^2_{\rm
  cr}$. Thus, the situation for $g^2 > g^2_{\rm cr}$ is as close as possible
to the situation studied in Sect.~\ref{sec:miransky}. The coupling can simply
be approximated by a constant. For $|\Theta| \ll 1$ the
gauge coupling runs very slowly ("walks") into its IR fixed point once it has
passed $g^2_{\rm cr}$. This walking behavior for $g^2 \gtrsim g^2_{\rm cr}$
then gives rise to sizable corrections to the exponential scaling
behavior.

Finally, we would like to discuss the finite-temperature many-flavor phase
boundary in QCD.  In~\cite{Braun:2005uj,Braun:2006jd} it was found that the
scaling of the phase boundary is consistent with the pure power-law scaling
behavior~\eqref{eq:TcrTheta}. From the above discussion this result is {now
understandable} since the exponential scaling behavior sets in only very close to
$\Nfcr$ for $\Nfcr\approx 12$ and thus remains invisible in numerical fits
over a wider range of $\Nf$. Of course, power-law scaling behavior for
the chiral phase-transition temperature still remains an upper bound even if
we took into account the exponential factor in Eq.~\eqref{eq:slawcorr}. This
is due to the fact that strong fluctuations of Goldstone modes in the IR may
yield further corrections and lower the phase transition temperature, see e.~g.
Ref.~\cite{Braun:2009si}.  Whether these corrections at finite temperature
yield additional corrections to the scaling behavior cannot be answered within
the scaling analysis presented in this work.  However, it may very well be that
such corrections depend only on $\Nf$ but not on $\Nfcr$. Nevertheless, we
would like to stress that a further investigation of the finite-temperature
scaling behavior at the {\it quantum critical point}, $\Nf=\Nfcr$, seems 
worthwhile in QCD since the scaling behavior in $\Nf$ direction may
significantly differ from the expected power-law scaling behavior in the
temperature direction at fixed $\Nf$, see~\cite{Pisarski:1983ms}.

\section{Quantitative scaling analysis in many-flavor QCD}\label{sec:manyflavor}
Having derived analytic scaling relations for physical observables in the previous sections, 
we present results from a numerical study of the scaling behavior in QCD with many flavors
in this section.
\subsection{Renormalization group setup}
Our numerical analysis is based on previous works on strongly-flavored gauge theories
in the framework of a functional RG approach using the Wetterich equation~\cite{Wetterich:1992yh}, 
for reviews, see \cite{Reuter:1996ub,Litim:1998nf,Bagnuls:2000ae,Berges:2000ew,Polonyi:2001se,Delamotte:2003dw,%
Pawlowski:2005xe,Gies:2006wv,Delamotte:2007pf,Rosten:2010vm}. 
In~\cite{Gies:2005as} the zero-temperature
quantum phase transition of QCD with $\Nc$ colors and $\Nf$ flavors has been computed using the
functional RG. The phase diagram at finite temperature as a function of $\Nf$ has first been computed 
in~\cite{Braun:2005uj,Braun:2006jd}. We  briefly review these results in this section
and employ them for our numerical analysis of the scaling behavior. 

In \cite{Gies:2005as,Braun:2005uj,Braun:2006jd} the RG flow of QCD starting from the
microscopic degrees of freedom in terms of quarks and gluons was studied
within a covariant derivative expansion. A crucial ingredient for chiral symmetry breaking are the
scale-dependent gluon-induced quark self-interactions of the type included 
in~\eqref{eq:miransky_ansatz}.
We note that dynamical quarks influence the RG flow of QCD by qualitatively
different mechanisms. First, quark fluctuations directly modify the running of 
the gauge coupling due to the screening nature of these fluctuations.
On the other hand, gluon exchange between quarks induces
quark self-interactions which can become relevant operators in the IR 
as we have already discussed in the previous sections. These two mechanisms
strongly influence each other as well. As we have seen, however, it is possible
to disentangle the system once we accept that
these fluctuations can be associated with different scales in the problem.

From now on we restrict ourselves to $d=4$ Euclidean space-time dimensions
and work solely in the Landau gauge. In a consistent and systematic operator
expansion of the effective action, the lowest nontrivial order is given by~\cite{Gies:2003dp}
\begin{widetext}
\begin{eqnarray}
\Gamma_k&=&
\int d^4x \Bigg\{ \bar{\psi}({\rm i}\partial\!\!\!\slash\; +
\bar{g}A\!\!\!\slash\;)\psi 
+\frac{1}{2} \Big[
  \bar\lambda_-(\text{V--A}) +\bar\lambda_+ (\text{V+A})
+\bar\lambda_\sigma (\text{S--P}) +\bar\lambda_{\text{VA}}
  [2(\text{V--A})^{\text{adj}}\!+({1}/{\Nc})(\text{V--A})] \Big]\Bigg\}.
\label{equ::truncation}
\end{eqnarray}
\end{widetext}
This ansatz for the effective action underlies our non-perturbative RG study.
The four-fermion interactions occurring here have been classified
according to their color and flavor structure. Color and flavor
singlets are
\begin{eqnarray}
(\text{V--A})&=&(\yb\gamma_\mu\psi)^2 + (\yb\gamma_\mu\gamma_5\psi)^2,
\\
(\text{V+A}) &=&(\yb\gamma_\mu\psi)^2 - (\yb\gamma_\mu\gamma_5\psi)^2 ,
\end{eqnarray}
where (fundamental) color ($i,j,\dots$) and flavor ($\chi,\xi,\dots$)
indices are contracted pairwise, e.g., $(\yb\psi)\equiv (\yb_i^{\chi}
\psi_i^{\chi})$.  The remaining operators have non-singlet color or flavor
structure,
\begin{eqnarray}
(\text{S--P})&=&\!(\yb^{\chi}\psi^{\xi})^2\!-(\yb^{\chi}\gamma_5\psi^{\xi})^2
\!\equiv\!
   (\yb_i^{\chi}\psi_i^{\xi})^2\!-(\yb_i^{\chi}\gamma_5\psi_i^{\xi})^2\!,\nonumber\\
(\text{V--A})^{\text{adj}}\!\!\!&=&\!(\yb \gamma_\mu T^a\psi)^2 
   + (\yb\gamma_\mu\gamma_5 T^a\psi)^2, \label{eq::colorflavor}
\end{eqnarray}
where $(\yb^{\chi}\psi^{\xi})^2\equiv \yb^{\chi}\psi^{\xi} 
\yb^{\xi} \psi^{\chi}$, etc., and $(T^a)_{ij}$ denote the generators of 
the gauge group in the fundamental representation.  

We stress that the set of fermionic self-interactions introduced in
Eq.~\eqref{equ::truncation} forms a complete basis. This means that any other
pointlike four-fermion interaction which is invariant under
\mbox{$\textrm{SU}(\Nc)$} gauge symmetry and
$\textrm{SU}(\Nf)_{\textrm{L}}\times \textrm{SU}(\Nf)_{\textrm{R}}$ flavor
symmetry can be related to those in \Eqref{equ::truncation} by means of Fierz
transformations. In our numerical analysis, we neglect
U${}_{\text{A}}(1)$-violating interactions induced by topologically
non-trivial gauge configurations, since we expect them to become relevant only
inside the \xsb\ regime or for small $\Nf$. In addition, the lowest-order
U${}_{\text{A}}$(1)-violating term schematically is $\sim
(\yb\psi)^{\Nf}$. Thus, larger $\Nf$ correspond to larger RG ``irrelevance''
by naive power-counting. Moreover, interactions of the type $\sim
(\yb\psi)^{\Nf}$ for $\Nf>3$ do not contribute directly to the flow of the four-fermion
interactions due to the one-loop structure of the underlying RG equation for
the effective action.

As a severe approximation, we drop any nontrivial momentum
dependencies of the $\bar\lambda$'s and study these couplings in the
point-like limit $\bar\lambda(|p_i|\ll k)$ in our scaling analysis.
Therefore our ansatz for the effective action does not allow us to 
study QCD properties in the chirally broken regime, since, e.~g., mesons
manifest themselves as momentum singularities in the
$\bar\lambda$'s. Nonetheless, our point-like approximation can be reasonable
in the chirally symmetric regime. This has been indeed shown in~\cite{Gies:2005as}, 
where the regularization-scheme independence of universal quantities 
has been found to hold remarkably well in the point-like limit.
 
Using the truncated effective action \eqref{equ::truncation}, we obtain 
the following $\beta$ functions for the dimensionless couplings 
$\lambda_i=\bar{\lambda_i}/k^2$ (see \cite{Gies:2003dp,Gies:2005as}):
\begin{widetext}
\begin{eqnarray}
\!\pat\lm
&=& 2\lm\!
    -4v_4 \lFBo\left[ \frac{3}{\Nc}g^2\lm
            -3g^2 \lva \right]
    \label{eq:lm}
-\frac{1}{8}v_{4}\lFB\left[\frac{12+9\Nc^2}{\Nc^2}g^{4} \right]
\\\nonumber
&&\quad\quad-8 v_4\lF \Big\{-\Nf\Nc(\lm^2+\lp^2) + \lm^2
\!\!-2(\Nc+\Nf)\lm\lva
       +\Nf\lp\lsf + 2\lva^2 \Big\},
%\\
\end{eqnarray}
\begin{eqnarray}
\!\pat\lp &=&2 \lp\! -4v_4\lFBo \left[
-\frac{3}{\Nc}g^2\lp\right]
    \label{eq:lp}
-\frac{1}{8}v_{4}\lFB\left[
    -\frac{12+3\Nc^2}{\Nc^2} g^{4} \right]
\\\nonumber
&&%\!\!\!\!\!
-8 v_4 \lF \Big\{ - 3\lp^2 - 2\Nc\Nf\lm\lp - 2\lp(\lm+(\Nc+\Nf)\lva)
        + \Nf\lm\lsf
+ \lva\lsf
        +\casel{1}{4}\lsf{}^2 \Big\},
    \nonumber%\\
\end{eqnarray}
\begin{eqnarray}
\!\pat\lsf
&=&
    2\lsf\! -4v_4 \lFBo \left[6\Cas\, g^2\lsf
    -6g^2\lp \right]\label{eq:lsf}
-\frac{1}{4} v_4 \lFB \Big[ -\frac{24
    -9\Nc^2}{\Nc}\, g^4  \Big] \\\nonumber
&&\quad\quad
 -8 v_4 \lF
     \! \Big\{\! 2\Nc \lsf^2\!  -\! 2\lm\lsf\! - 2\Nf\lsf\lva\!\!
- \!6\lp\lsf\! \Big\},
        \nonumber%\\
\end{eqnarray}
\begin{eqnarray}
\!\pat\lva
&=& 2 \lva\!-4v_4 \lFBo \left[
\frac{3}{\Nc}g^2\lva -3g^2\lm \right]
    \label{eq:lva} 
-\frac{1}{8} v_4 \lFB \left[ -\frac{24 - 3\Nc^2}{\Nc}
g^4 \right]
    \\
&&\quad\quad -8 v_4 \lF\!
  \Big\{\! - \!(\Nc+\Nf)\lva^2\! + 4\lm\lva\!
        - \casel{1}{4} \Nf \lsf^2\Big\}.\nonumber
\end{eqnarray}
\end{widetext}
Here, $\Cas=(\Nc^2-1)/(2\Nc)$ is a Casimir operator of the gauge group,
and $v_4=1/(32\pi^2)$. The regularization-scheme dependence of the RG flow
equations is controlled by (dimensionless) threshold functions $l$ which arise
from Feynman diagrams and incorporate fermionic and/or bosonic fields~\cite{Jungnickel:1995fp}. 
For the optimized regulator~\cite{Litim:2000ci,Litim:2001up,Litim:2001fd}, we find

\begin{equation}
\label{optimized}
\lF=\frac{1}{2}\,,\quad
\lFBo=1-\frac{\eta_{\rm A}}{6}\,, 
\quad\lFB=\frac{3}{2}-\frac{\eta_{\rm A}}{6}\,.
\end{equation}
In our numerical analysis, we have dropped contributions from the anomalous
dimensions of the fermions and the gauge coupling $\eta_A=\beta_{g^2}/g^2$.
While the first one is proportional to the gauge-fixing parameter and vanishes
identical in the Landau gauge in the chirally symmetric regime~\cite{Gies:2003dp},
we have found by a comparison of our numerical results with those
from~\cite{Gies:2005as} that the contributions $\propto \eta_A$ in the
threshold function do not strongly affect our result for $\Nfcr$. {In fact, we
  have $\eta_{A}\to 0$ for $\Nf\to N_{\rm f}^{\rm a.f.}$ and $g^2<
  g^2_{\ast}$. Moreover, we find for $g^2< g^2_{\ast}$ that
  $|\eta_{A}^{\rm 2-loop}| \lesssim 1$ for $\Nf \gtrsim 11$ and
 $|\eta_{A}^{\rm 4-loop}| \lesssim 0.5$ for $\Nf \gtrsim 8$. In
  total, this may lead to quantitative corrections at most on the percent level.

Let us now discuss the running of the gauge coupling. Even though the running
coupling has been computed within the functional RG
{approach~\cite{Reuter:1993kw,Gies:2002af,Pawlowski:2003hq,Braun:2005uj,Braun:2006jd}}, we
employ for simplicity the two- and four-loop result obtained in the $\MSbar$
scheme \cite{vanRitbergen:1997va,Czakon:2004bu}, as our results show a
satisfactory convergence in the strongly flavored regime. We will often
restrict ourselves to the two-loop case, as it already shows all qualitative features
and can be dealt with analytically. The analytic expression for the two-loop
$\beta_{g^2}$ function reads explicitly:
\be
\partial_t g^2 = -\left(\beta_0 + \beta_1\left(\frac{g^2}{16\pi^2}\right) +\dots
\right)\frac{g^2}{8\pi^2}\,,
\ee
with
\be
\beta_{0}=\frac{11}{3}\Nc-\frac{2}{3}\Nf,\,\,
\beta_{1}=\frac{34\Nc^3+3\Nf-13\Nc^2\Nf}{3\Nc}\,.
\ee
Note that the chosen regularization scheme in the matter sector and the
$\MSbar$ scheme do not coincide. This inconsistency results in
an error for our estimate for the critical number of quark flavors. Since we
are rather interested in the scaling behavior which is related to the
universal critical exponent $\Theta$, our results are only influenced
indirectly by this approximation\footnote{Of course, the actual value of
  $\Theta_0=\Theta(\Nfcr)$ depends on the actual value of $\Nfcr$ which
  itself, as a universal quantity, depends on the difference of the
  scheme-dependent quantities $g^2_{\rm cr}$ and $g^2_{\ast}$.}. Due to
  this scheme dependence, the results using the four-loop running may not
  necessarily be considered as a more precise calculation. Instead, the
  difference between two-loop and four-loop $\MSbar$ results should be viewed
  as an estimate of the dependence of our results on the quantitative details
  of the running gauge sector.

A comment on contributions to the running of the gauge coupling induced by the
presence of the quark self-interactions $\lambda_i$ is in order here: To
render the RG flow gauge invariant we have to take regulator-dependent
Ward-Takahashi identities into
account~\cite{Ellwanger:1994iz,Reuter:1993kw}. In the present case, these
symmetry constraints yield contributions to the running of the gauge coupling
which depend on the quark self-interactions. However, these contributions are
proportional to the $\beta$ functions of the four-fermion couplings, as has
been pointed out in Ref.~\cite{Gies:2003dp,Gies:2005as}. Therefore, these
contributions vanish as long as the four-fermion couplings are at their fixed points, i.~e. as long as $g^2 \leq g^2_{\rm cr}$.  Thus, we
expect that these contributions do not alter the
scaling-law~\eqref{eq:ksbscaling} in leading order\footnote{In addition to
 the presented next-to-leading order corrections to~\eqref{eq:ksbscaling},
  these symmetry constraints may shift the fixed-point value $g^2_{\ast}$ of
  the gauge coupling and therefore cause additional higher-order corrections
  to the scaling law~\eqref{eq:ksbscaling}.}.  In particular, the power-law
behavior is unaffected by these corrections arising due to symmetry
constraints. In the present approximation, we ignore these
corrections in our numerical analysis.

\subsection{Miransky-type scaling}
\begin{figure}[t]
\includegraphics[scale=0.65]{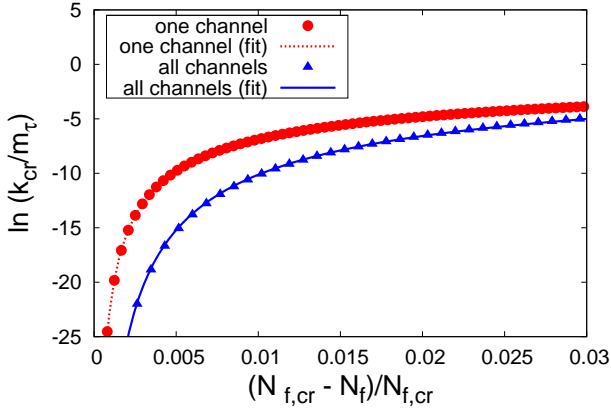}
\caption{
Logarithm of the (chiral) symmetry-breaking scale $\ln (\ksb/m_{\tau})$ as a function of $(\Nfcr-\Nfcr)/\Nfcr$ 
for an $\Nf$-dependent but scale-independent, i.e. constant gauge
coupling. {The corresponding fits are given in \Eqref{eq:Mirfits}.}  
}
\label{fig:miransky}
\end{figure}

Let us start with a numerical analysis of many-flavor QCD with a constant gauge coupling:
\be
\partial_t g^2=0\,.\nn
\ee
As discussed above, the gauge coupling can then be considered as an "external" $\Nf$-dependent
parameter of the theory. For our numerical study we choose the fixed-point value of the
gauge-coupling at the two-loop level:
\be
g^2_{\ast,{\rm 2-loop}}(\Nf)=\frac{16 (11 \Nc^2 - 2 \Nc \Nf) \pi^2}{  13 \Nc^2 \Nf - 34 \Nc^3 - 3 \Nf}
\,.
\ee

In the matter sector we employ two different truncations to which we refer as
{\it one-channel} and {\it all-channels} approximation. The latter one is Fierz
  complete.
In the all-channels approximation we take into account the full set of flow equations~\eqref{eq:lm}-\eqref{eq:lva}, 
while we only take into account the RG flow of the scalar-pseudoscalar channel $\lambda_{\sigma}$
in the one-channel approximation and set all other four-fermion couplings to zero:
\be
\!\pat\lsf
= 2\lsf\! 
- a_{\sigma}\lsf^2 - b_{\sigma}\lsf g^2 - c_{\sigma} g^4\,,
\label{eq:lsf1channel}
\ee
with
\be
a_{\sigma}&=&\frac{\Nc}{4\pi^2}\,,\quad b_{\sigma}=\frac{3}{4\pi^2}\Cas\,,\nn\\
c_{\sigma}&=&\frac{3}{256\pi^2}\left(\frac{9\Nc^2 -24}{\Nc} \right)\,.
\ee
Here, we have adopted the conventions of Sect.~\ref{sec:miransky} for the coefficients $a,b,c$.

Now we can compute the critical values of the gauge coupling in the one- and
in the all-channels approximation. In the one-channel approximation we find
\begin{equation}
g^2_{\rm cr,one}\! = \frac{32\pi^2\left(2\Nc^3\! -2\Nc\! -\sqrt{3\Nc^6 - 8\Nc^4} \right)}
{3(4+\Nc^4)}\!\stackrel{(\Nc=3)}{\approx}\!10.86\,,
\end{equation}
which does not depend on $\Nf$. In the all-channels approximation the critical value has to
be computed numerically. As found in~\cite{Gies:2005as}, the resulting critical value 
$g^2_{\rm cr,all}$ of the gauge coupling then depends on $\Nf$; for a given number of colors,
$g^2_{\rm cr,all}$ decreases weakly with increasing~$\Nf$.

The fixed-point value $g^2_{\ast,{\rm 2-loop}}$ together with the critical value of the gauge coupling 
can be used to estimate the critical
number of quark flavors above which there is no chiral symmetry breaking in the IR. In agreement
with the results given in Ref.~\cite{Gies:2005as}, we find
\begin{widetext}
\be
\Nfcr^{\rm one}= \frac{169 \Nc^6 \! - \! 136 \Nc^4 \! +\! 132 \Nc^2 \! - \! 68
   \sqrt{\Nc^4 \left(3 \Nc^2\! -\! 8\right)} \Nc^3}{58
 \Nc^5\!-\! 64 \Nc^3\! -\! 26 \sqrt{\Nc^4 \left(3
  \Nc^2\! -\! 8\right)} \Nc^2\! +\! 6 \sqrt{\Nc^4 \left(3
  \Nc^2\! -\! 8\right)}\! +\! 36 \Nc}
\stackrel{(\Nc =3)}{\approx} 11.7\,,\label{eq:Nfcr1channel}
\ee
\end{widetext}
for the one-channel approximation and
\be
\Nfcr^{\rm all}\approx 11.9
\ee
for the all-channels approximation. We may use our estimate for $\Nfcr$ from the 
one-channel approximation to estimate $\Nfcr$ in the limit $\Nc\to\infty$:
\be
\frac{\Nfcr^{\rm one}}{\Nc}=\frac{68 \sqrt{3}-169}{2 \left(13 \sqrt{3}-29\right)}\approx 3.95\,.
\ee
Our results for $\Nfcr$ are in accordance with the results from Dyson-Schwinger equations
in the rainbow-ladder approximation, see e.~g.~\cite{Appelquist:1998rb,Dietrich:2006cm,Fukano:2010yv}
as well as with those from current lattice simulations~\cite{Kogut:1982fn,Gavai:1985wi,Fukugita:1987mb,Brown:1992fz,Damgaard:1997ut,
  Iwasaki:2003de,Catterall:2007yx,Appelquist:2007hu,Deuzeman:2008sc,Deuzeman:2009mh,Appelquist:2009ty,
  Fodor:2009wk,Fodor:2009ff,Pallante:2009hu}.

Let us now study the dependence of the symmetry breaking scale $\ksb$ on $\Nf$
for the specific case $\Nc =3$. 
In Fig.~\ref{fig:miransky} we show our results for $\ln(\ksb/\Lambda)$ as function of 
$(\Nfcr-\Nf)/\Nf$ as obtained from the one-channel (dots) and from the 
all-channels (triangles) 
approximation using $g^2_{\ast,{\rm 2-loop}}$ as a fixed input parameter. 
As initial conditions for the $\lambda_i$'s for a given $g^2_{\ast,{\rm 2-loop}}(\Nf)$
we have used the solution of the coupled set of linear equations
\be
\frac{\partial ( \partial_t\lambda_i)}{\partial \lambda_i}=0\,,
\ee
where $i\in\{+,-,\sigma,{\rm VA}\}$. This corresponds to starting the flow at the maxima
(extrema) of the parabolas.

We observe that for a given $\Nf$ the symmetry breaking scale $\ksb$ is smaller in the all-channels
approximation compared to the one-channel approximation. The fits to the data points are 
also shown in Fig.~\ref{fig:miransky}. In agreement with our analytic results we find:
\be
\ln\ksb^{\rm one}&\approx& {\rm const.} - \frac{2.481}{|\Nfcr-\Nf|^{0.494}}\,,\nn\\
\ln\ksb^{\rm all}&\approx& {\rm const.} - \frac{3.932}{|\Nfcr-\Nf|^{0.516}}\,.\label{eq:Mirfits}
\ee
Thus, we clearly observe the expected exponential scaling behavior in the one-cannel and in the
all-channels approximation for $\Nf\to\Nfcr$. 

The result from the one-channel approximation is in reasonable agreement with the analytic leading-order (LO)
result found in Sect.~\ref{sec:miransky}:
\be
\ln\ksb^{\rm LO}&=&{\rm const.} - \frac{\pi}{2\epsilon\sqrt{|\alpha_1|
    |\Nfcr-\Nf|}} \nn\\
&\approx& {\rm const.} - \frac{2.386}{\sqrt{|\Nfcr-\Nf|}}\,. 
\ee
Note that $|\alpha_2/\alpha_1|\approx 0.273$. Differences to the numerical results are due to 
numerical errors of the fit and higher-order corrections which we have derived
in Sect.~\ref{sec:miransky}. 

\subsection{Power-law scaling and beyond}
\begin{figure}[t!]
\includegraphics[scale=0.6]{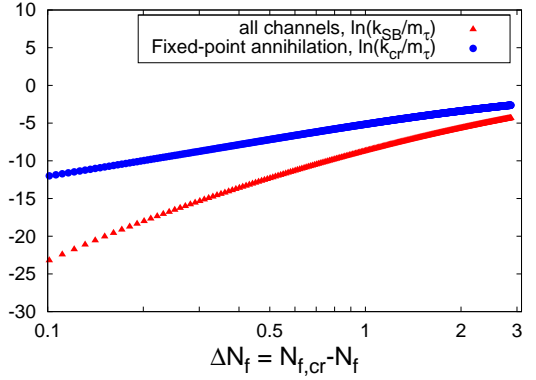}
\caption{$\Nf$ dependence of $\kcr$ and $\ksb$ as obtained from a study with a
  running coupling in the two-loop approximation. The criticality scale
    $\kcr$ (blue circles) is dominated by power-law scaling (straight line
    with slope $\sim|\Theta_0|^{-1}$ in
    this double-log plot), and clearly serves as an upper bound for the
    symmetry breaking scale $\ksb$ (red triangles), being a superposition of
    power-law and Miransky scaling. If the theories are probed at integer
    $\Nf$, i.e., $\Delta \Nf\gtrsim \mathcal{O}(1)$, the contribution due to
    Miransky scaling may not be visibile. A pure powerlaw fit to chiral
    observables $\sim \ksb$, may however overestimate the critical exponent $\Theta_0$. 
}
\label{fig:2loop}
\end{figure}

Let us now take into account the (momentum) scale-dependence of the running gauge coupling. 
In order to compare the theories with different flavor numbers we fix the scales by keeping the running 
coupling at the $\tau$-mass scale $\Lambda=m_{\tau}$ fixed to $\alpha(m_{\tau})\approx 0.322$.
Since we apply the truncation~\eqref{equ::truncation} to QCD, we do not consider the four-fermion 
couplings $\bar\lambda$ as independent external parameters as, e.g., in Nambu--Jona-Lasinio-type models. 
More precisely, we impose the boundary condition $\bar\lambda_i\to 0$ for
$k\to \infty$ which 
guarantees that the $\bar\lambda$'s at finite $k$ are solely generated by quark-gluon dynamics, 
e.g., by 1PI ``box'' diagrams with 2-gluon exchange, cf. Fig.~\ref{fig:feynman}(c). 

In Fig.~\ref{fig:2loop} and Fig.~\ref{fig:4loop} we show our results for the
  $\Nf$ dependence of the scales $\kcr$ and $\ksb$ as obtained from a study
  with a running gauge coupling in the two-loop and the four-loop
  approximation, respectively. Note that $\Nfcr$ becomes smaller when we employ the 
  running coupling in the four-loop approximation. We find 
\be
\Nfcr^{\rm 4-loop}\approx 10.0
\ee
in the all-channels approximation and 
$\Nfcr^{\rm 4-loop}\approx 9.8$ in the one-channel approximation, 
in agreement with Ref.~\cite{Gies:2005as}.

\begin{figure}[t!]
\includegraphics[scale=0.6]{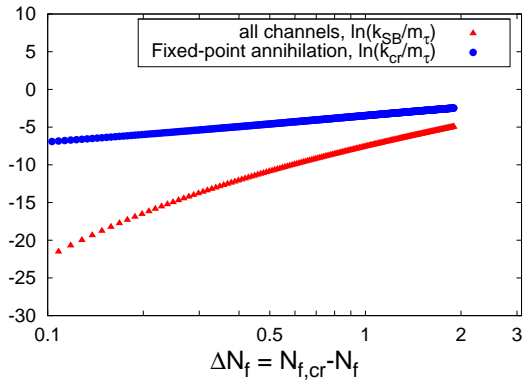}
\caption{$\Nf$ dependence of $\kcr$ and $\ksb$ as obtained from a study with a
  running gauge coupling in the four-loop approximation. The 
    contributions due to Miransky scaling, roughly parameterized by the
    difference between $\kcr$ (blue circles) and $\ksb$ (red triangles),
    extend to larger values of $\Delta \Nf=\Nfcr -\Nf$, as the estimate for
    the critical exponent $\Theta_0=\Theta(\Nfcr)$ at four loop is larger than
  at two-loop. In this perturbative estimate for the running coupling, the
  curves cannot be extended to larger values of $\Delta\Nf$, see text.}
\label{fig:4loop}
\end{figure}
\begin{figure*}[t]
\hspace*{0.0cm}
\includegraphics[scale=0.6]{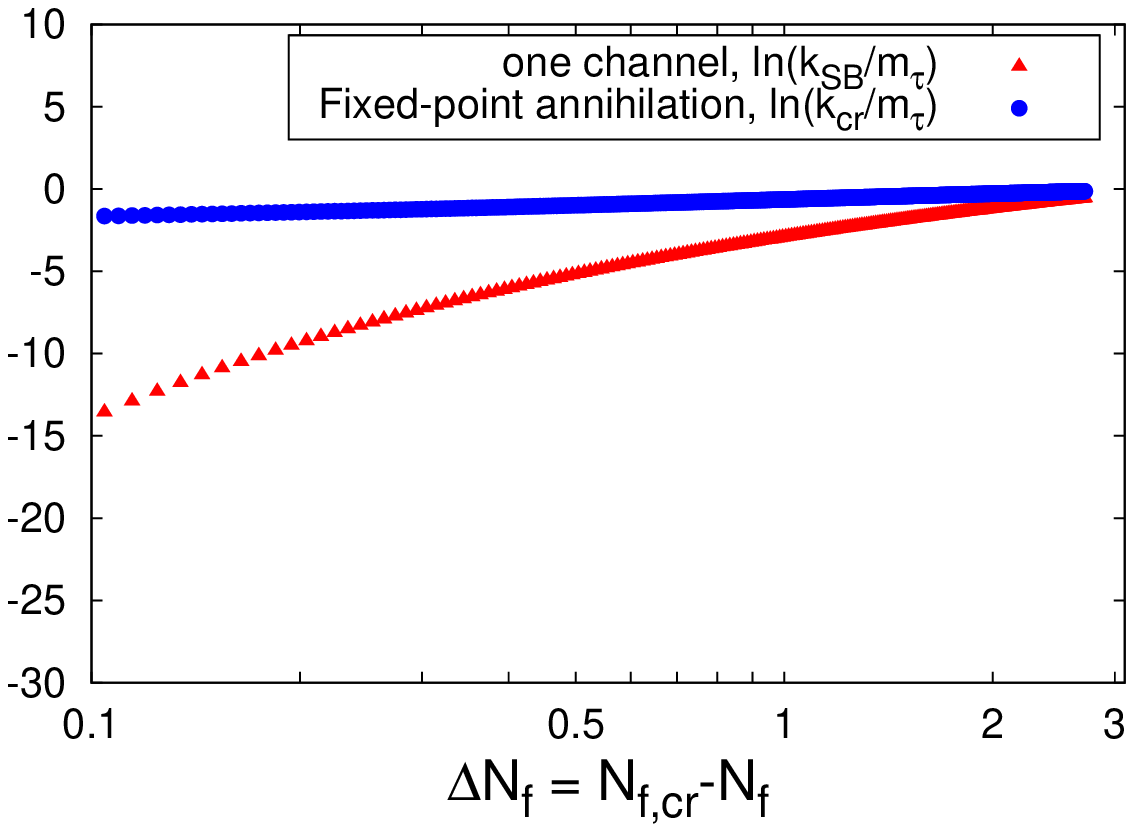}
\hspace*{1.5cm}
\includegraphics[scale=0.6]{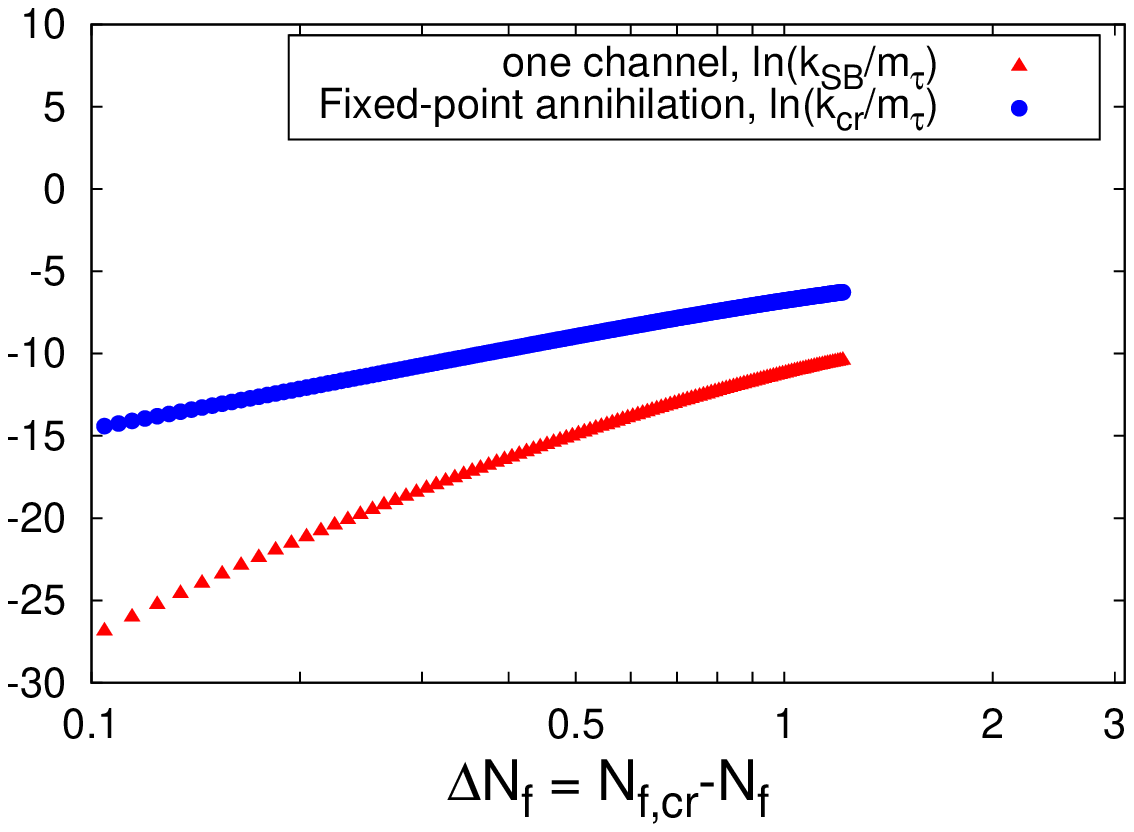}
\caption{$\Nf$ dependence of $\kcr$ (blue circles) and $\ksb$
  (red triangles) as obtained from a study with a model for the
    running gauge coupling, cf. \Eqref{eq:art2loop}, that allows to tune the
    critical exponent $\Theta$ by hand.  We show the results for
  $\Theta_0=|\Theta(\Nfcr)|\approx 4.3$ (left panel) and $|\Theta_0|\approx
  0.3$ (right panel). Contributions due to Miransky scaling are visible
    as deviations from a straight-line behavior (power law) in this double-log
    plot. These results confirm our estimate that the Miransky-scaling window is
    larger for larger $|\Theta_0|$ (left panel), whereas power-law scaling
    dominates for small $|\Theta_0|$ (right panel).}
\label{fig:thetadep}
\end{figure*}

The data points can be fitted to our analytic results for the scaling behavior of $\ksb$ and $\kcr$. For the 
all-channels approximation, we find
\be
\ln \kcr^{\rm 2-loop} &\approx& {\rm const.} + 2.566\, |\Nf-\Nfcr|\,,\\
\ln \ksb^{\rm 2-loop} &\approx& {\rm const.} -
\frac{3.401}{|\Nf-\Nfcr|^{0.54}} \nn\\
&&+ 2.540 \ln |\Nf-\Nfcr|\,,\label{eq:2loopfit}
\ee
and
\be
\ln \kcr^{\rm 4-loop} &\approx& {\rm const.} + 1.180\, |\Nf-\Nfcr|\,,\\
\ln \ksb^{\rm 4-loop} &\approx& {\rm const.} -
\frac{5.196}{|\Nf-\Nfcr|^{0.52}}\nn\\
&& + 1.171 \ln |\Nf-\Nfcr|\,.\label{eq:4loopfit}
\ee
Thus, the fits are in reasonable agreement with our analytic predictions. For
the multi-parameter fits~\eqref{eq:2loopfit} and~\eqref{eq:4loopfit}, we have
fixed the coefficient of the $\ln$-term which is the inverse critical exponent
$\Theta_0=\Theta(\Nfcr)$.  It should be stressed that the predicted values for
the critical exponent $\Theta(\Nfcr)$ are substantially different for the
running coupling in the two- and four-loop approximation
(cf. Fig~\ref{fig:Theta}), 
\be
\frac{1}{|\Theta(\Nfcr)|}&\approx& 2.540\quad \text{(two-loop)}\,, \nn\\
\frac{1}{|\Theta(\Nfcr)|}&\approx& 1.171\quad \text{(four-loop)}\,.
\ee
In Fig.~\ref{fig:2loop} and Fig.~\ref{fig:4loop} we observe that the critical exponent $\Theta$ clearly influences
the scaling behavior close to the quantum critical point $\Nfcr$. In agreement with our analytic findings,
the size of the regime with exponential scaling increases with increasing critical exponent~$\Theta$. Using Eq.~\eqref{eq:size}
we can give a quantitative estimate for the size of the regime in which the
exponential scaling behavior dominates. For the one-channel approximation, cf. Eq.~\eqref{eq:Nfcr1channel}, we find:
\be
\Delta N_f := |\Nf-\Nfcr| \lesssim 0.3 \;\;(\text{two-loop}, \Nfcr\approx 11.7).
\ee
Using a running coupling in four-loop order ($\Nfcr\approx 9.8$) the size of
this Miransky scaling regime can be estimated to be larger than one
flavor. This is in agreement with our numerical results, see
Figs.~\ref{fig:2loop} and ~\ref{fig:4loop}. With this perturbative
  estimate for the running coupling, however, the curves in Figs.~\ref{fig:2loop} and
  ~\ref{fig:4loop} cannot be extended to larger values of $\Delta\Nf=
  \Nfcr-\Nf$.  For instance, in the four-loop case, we have $\Nfcr\simeq9.8$ on the
  one hand. On the other hand, the Caswell-Banks-Zaks fixed point vanishes for
  $\Nf\lesssim 8$. Our RG arguments based on expansions about an IR fixed
  point hence only extend to $\Delta N_{{\rm f},{\text{max}}} \simeq 1.8$,
  cf. Fig.~\ref{fig:4loop}. In nonperturbative functional studies where an IR
  fixed point appears to exist already in the pure gauge sector and thus also
  at lower $\Nf$ \cite{vonSmekal:1997is,vonSmekal:1997vx,Lerche:2002ep,Alkofer:2004it,Fischer:2008uz,
  Fischer:2009tn,Fischer:2006vf,Aguilar:2009nf,Braun:2005uj,Braun:2006jd}, no restriction on
  $\Delta \Nf$ arises.

To summarize: since $\Nfcr\gtrsim 9$ in current lattice
simulations~\cite{Kogut:1982fn,Gavai:1985wi,Fukugita:1987mb,Brown:1992fz,Damgaard:1997ut,
  Iwasaki:2003de,Catterall:2007yx,Appelquist:2007hu,Deuzeman:2008sc,Deuzeman:2009mh,Appelquist:2009ty,
  Fodor:2009wk,Fodor:2009ff,Pallante:2009hu}, we expect that the pure
exponential scaling behavior is difficult to resolve and the corrections due
to the running of the gauge coupling ($\Theta$) might be more relevant for
lattice simulations. From the viewpoint of such simulations, one might be
interested in keeping the power of the "Miransky" term fixed to $1/2$ and use
the scaling law to fit $\Nfcr$ and the {critical exponent
  $\Theta_0=\Theta(\Nfcr)$.}

To illustrate the influence of the critical exponent $\Theta$ we have also
computed the scaling behavior of the scales $\kcr$ and $\ksb$ using a
model for the running gauge coupling. This model is inspired by
the two-loop approximation modified by an artificial higher-order
term. The latter is constructed such that the critical exponent $\Theta$ can be changed by
  hand, still leaving the two-loop fixed point unaffected:
\be
\partial_t g^2 \equiv\beta_{g^2} = \beta_{g^2}^{\rm 2-loop} + \phi\, g^6(g^2 - g^2_{\ast,\rm 2-loop})\,,
\label{eq:art2loop}
\ee
where the parameter $\phi$ allows us to change $\Theta$ without changing $\Nfcr$. 
In Fig.~\ref{fig:thetadep} we present our results $\ksb$ and $\kcr$ for $\phi=0.003$ (i.~e. $|\Theta(\Nfcr)|\approx 4.3$)
in the left panel and for $\phi=-0.0001$ (i.~e. $|\Theta(\Nfcr)|\approx 0.3$) in the right panel. The results clearly confirm
that the size of the exponential-scaling regime depends strongly on $\Theta$.

\section{Conclusions}
In this work we have analyzed how physical observables in asymptotically free gauge theories, such as QCD 
or QED${}_3$, scale when the number of  flavors is varied. When the number $\Nf$ of fermion flavors
in such theories is increased, a regime may open up along the $\Nf$ axis 
in which the theory is asymptotically free but remains chirally symmetric in the infrared. This gives rise to the
existence of a quantum critical point on the $\Nf$ axis. The exact determination of the location of this
quantum critical point in QCD as well as in QED${}_3$ is currently a very active frontier in theoretical physics.

Even though we have presented estimates for $\Nfcr$ as obtained from a functional RG approach,
see also~\cite{Gies:2005as,Braun:2006jd}, the focus of this work is on the actual scaling behavior 
of physical observables close to the quantum critical point. This scaling
behavior of observables such as the fermion condensate close to $\Nfcr$ is not only interesting in its
own right but may also help to guide future lattice simulations in this field.

Ignoring the running of the gauge coupling, it has long been known that
physical observables obey Miransky scaling, i.e., an exponential scaling law
close to the quantum critical point~\cite{Berezinskii,Berezinskii2,Kosterlitz:1973xp,Miransky:1988gk,Miransky:1996pd}. 
In more recent studies~\cite{Braun:2006jd,Braun:2009ns} we have shown that an upper
bound for the scaling behavior of, e.~g. the chiral phase transition
temperature, can be derived from an analysis of the fixed-point structure in
the matter sector. In combination with the running of the gauge coupling in
its fixed-point regime, the upper bound for physical observables then scales
according to a power law. The associated critical exponent is related to the
universal critical exponent $\Theta$ at the IR fixed-point of the gauge
coupling, see Eq.~\eqref{eq:kcr}.

From another viewpoint, we have shown that Miransky scaling close to $\Nfcr$
receives universal power-law corrections, which are uniquely determined by the
critical exponent at the IR fixed-point of the gauge coupling. Both scaling
laws follow from one and the same set of RG flow equations and can be
considered as two different limits of a {general scaling law}: pure exponential
Miransky scaling arises in the limit of large $\Theta\to \infty$, whereas
power-law scaling becomes more prominent at small $\Theta\to 0$.
Quantitatively, we have estimated the size of the regime with almost pure
exponential scaling in strongly-flavored QCD and found it to be small,
$|\Nfcr-\Nf|\lesssim 0.3$ for $\Nfcr\approx 11.7$.  Outside this regime the
scaling behavior of physical observables is controlled by the critical
exponent $\Theta$. Our numerical analysis of scaling in many-flavor QCD based
on functional RG methods is indeed in agreement with these analytic findings.

Finally we would like to add that the scaling behavior close to the quantum
critical point can be contaminated by the scale-fixing procedure. As also
argued in~\cite{Braun:2009ns}, a comparison of theories with different $\Nf$
is not unique for non-conformal theories but indeed requires a specific choice
of a dimensionful scale. This scale is used as a ruler for the different
theories. In the present work, we have fixed the scales by keeping the gauge
coupling fixed to the same value for all $\Nf$ at an initial mid-momentum
scale, e.~g. the $\tau$ mass scale in QCD. Alternatively, theories with
different $\Nf$ can be fixed by keeping an IR observable characteristic for
the ordered phase for all $\Nf$ fixed,
{say the pion decay constant $f_\pi$ or $T_{\rm cr}$} in QCD. However, the behavior of physical observables at
the quantum critical point is then discontinuous.

In any case, the scaling relation~\eqref{eq:slawcorr} is a parameter-free testable prediction for the
behavior of physical observables near the quantum critical point. On the one hand, it might be 
tested directly by lattice simulations. On the other hand, our prediction might also be helpful to guide
future lattice simulations of strongly-flavored gauge theories.

\acknowledgments The authors thank D. D. Dietrich, J. M. Pawlowski and
F. Sannino for useful discussions and acknowledge {support by the DFG under
  grants Gi~328/1-4, Gi~328/5-1 (Heisenberg program), FOR 723 and GRK 1523/1}
and by the Helmholtz-University Young Investigator Grant No.~VH-NG-332 and by
the Helmholtz International Center for FAIR within the LOEWE program of the
State of Hesse.

\bibliography{references}

\begin{thebibliography}{105}
\expandafter\ifx\csname natexlab\endcsname\relax\def\natexlab#1{#1}\fi
\expandafter\ifx\csname bibnamefont\endcsname\relax
  \def\bibnamefont#1{#1}\fi
\expandafter\ifx\csname bibfnamefont\endcsname\relax
  \def\bibfnamefont#1{#1}\fi
\expandafter\ifx\csname citenamefont\endcsname\relax
  \def\citenamefont#1{#1}\fi
\expandafter\ifx\csname url\endcsname\relax
  \def\url#1{\texttt{#1}}\fi
\expandafter\ifx\csname urlprefix\endcsname\relax\def\urlprefix{URL }\fi
\providecommand{\bibinfo}[2]{#2}
\providecommand{\eprint}[2][]{\url{#2}}

\bibitem[{\citenamefont{Weinberg}(1979)}]{Weinberg:1979bn}
\bibinfo{author}{\bibfnamefont{S.}~\bibnamefont{Weinberg}},
  \bibinfo{journal}{Phys. Rev.} \textbf{\bibinfo{volume}{D19}},
  \bibinfo{pages}{1277} (\bibinfo{year}{1979}).

\bibitem[{\citenamefont{Holdom}(1981)}]{Holdom:1981rm}
\bibinfo{author}{\bibfnamefont{B.}~\bibnamefont{Holdom}},
  \bibinfo{journal}{Phys. Rev.} \textbf{\bibinfo{volume}{D24}},
  \bibinfo{pages}{1441} (\bibinfo{year}{1981}).

\bibitem[{\citenamefont{Hong et~al.}(2004)\citenamefont{Hong, Hsu, and
  Sannino}}]{Hong:2004td}
\bibinfo{author}{\bibfnamefont{D.~K.} \bibnamefont{Hong}},
  \bibinfo{author}{\bibfnamefont{S.~D.~H.} \bibnamefont{Hsu}},
  \bibnamefont{and} \bibinfo{author}{\bibfnamefont{F.}~\bibnamefont{Sannino}},
  \bibinfo{journal}{Phys. Lett.} \textbf{\bibinfo{volume}{B597}},
  \bibinfo{pages}{89} (\bibinfo{year}{2004}), \eprint{hep-ph/0406200}.

\bibitem[{\citenamefont{Sannino and Tuominen}(2005)}]{Sannino:2004qp}
\bibinfo{author}{\bibfnamefont{F.}~\bibnamefont{Sannino}} \bibnamefont{and}
  \bibinfo{author}{\bibfnamefont{K.}~\bibnamefont{Tuominen}},
  \bibinfo{journal}{Phys. Rev.} \textbf{\bibinfo{volume}{D71}},
  \bibinfo{pages}{051901} (\bibinfo{year}{2005}), \eprint{hep-ph/0405209}.

\bibitem[{\citenamefont{Dietrich et~al.}(2005)\citenamefont{Dietrich, Sannino,
  and Tuominen}}]{Dietrich:2005jn}
\bibinfo{author}{\bibfnamefont{D.~D.} \bibnamefont{Dietrich}},
  \bibinfo{author}{\bibfnamefont{F.}~\bibnamefont{Sannino}}, \bibnamefont{and}
  \bibinfo{author}{\bibfnamefont{K.}~\bibnamefont{Tuominen}},
  \bibinfo{journal}{Phys. Rev.} \textbf{\bibinfo{volume}{D72}},
  \bibinfo{pages}{055001} (\bibinfo{year}{2005}), \eprint{hep-ph/0505059}.

\bibitem[{\citenamefont{Dietrich and Sannino}(2007)}]{Dietrich:2006cm}
\bibinfo{author}{\bibfnamefont{D.~D.} \bibnamefont{Dietrich}} \bibnamefont{and}
  \bibinfo{author}{\bibfnamefont{F.}~\bibnamefont{Sannino}},
  \bibinfo{journal}{Phys. Rev.} \textbf{\bibinfo{volume}{D75}},
  \bibinfo{pages}{085018} (\bibinfo{year}{2007}), \eprint{hep-ph/0611341}.

\bibitem[{\citenamefont{Ryttov and Sannino}(2007)}]{Ryttov:2007sr}
\bibinfo{author}{\bibfnamefont{T.~A.} \bibnamefont{Ryttov}} \bibnamefont{and}
  \bibinfo{author}{\bibfnamefont{F.}~\bibnamefont{Sannino}},
  \bibinfo{journal}{Phys. Rev.} \textbf{\bibinfo{volume}{D76}},
  \bibinfo{pages}{105004} (\bibinfo{year}{2007}), \eprint{0707.3166}.

\bibitem[{\citenamefont{Antipin and Tuominen}(2010)}]{Antipin:2009wr}
\bibinfo{author}{\bibfnamefont{O.}~\bibnamefont{Antipin}} \bibnamefont{and}
  \bibinfo{author}{\bibfnamefont{K.}~\bibnamefont{Tuominen}},
  \bibinfo{journal}{Phys. Rev.} \textbf{\bibinfo{volume}{D81}},
  \bibinfo{pages}{076011} (\bibinfo{year}{2010}), \eprint{0909.4879}.

\bibitem[{\citenamefont{Sannino}(2009{\natexlab{a}})}]{Sannino:2009za}
\bibinfo{author}{\bibfnamefont{F.}~\bibnamefont{Sannino}},
  \bibinfo{journal}{Acta Phys. Polon.} \textbf{\bibinfo{volume}{B40}},
  \bibinfo{pages}{3533} (\bibinfo{year}{2009}{\natexlab{a}}),
  \eprint{0911.0931}.

\bibitem[{\citenamefont{Caswell}(1974)}]{Caswell:1974gg}
\bibinfo{author}{\bibfnamefont{W.~E.} \bibnamefont{Caswell}},
  \bibinfo{journal}{Phys. Rev. Lett.} \textbf{\bibinfo{volume}{33}},
  \bibinfo{pages}{244} (\bibinfo{year}{1974}).

\bibitem[{\citenamefont{Banks and Zaks}(1982)}]{Banks:1981nn}
\bibinfo{author}{\bibfnamefont{T.}~\bibnamefont{Banks}} \bibnamefont{and}
  \bibinfo{author}{\bibfnamefont{A.}~\bibnamefont{Zaks}},
  \bibinfo{journal}{Nucl. Phys.} \textbf{\bibinfo{volume}{B196}},
  \bibinfo{pages}{189} (\bibinfo{year}{1982}).

\bibitem[{\citenamefont{Pisarski}(1984)}]{Pisarski:1984dj}
\bibinfo{author}{\bibfnamefont{R.~D.} \bibnamefont{Pisarski}},
  \bibinfo{journal}{Phys. Rev.} \textbf{\bibinfo{volume}{D29}},
  \bibinfo{pages}{2423} (\bibinfo{year}{1984}).

\bibitem[{\citenamefont{Appelquist et~al.}(1986)\citenamefont{Appelquist,
  Bowick, Karabali, and Wijewardhana}}]{Appelquist:1986fd}
\bibinfo{author}{\bibfnamefont{T.~W.} \bibnamefont{Appelquist}},
  \bibinfo{author}{\bibfnamefont{M.~J.} \bibnamefont{Bowick}},
  \bibinfo{author}{\bibfnamefont{D.}~\bibnamefont{Karabali}}, \bibnamefont{and}
  \bibinfo{author}{\bibfnamefont{L.~C.~R.} \bibnamefont{Wijewardhana}},
  \bibinfo{journal}{Phys. Rev.} \textbf{\bibinfo{volume}{D33}},
  \bibinfo{pages}{3704} (\bibinfo{year}{1986}).

\bibitem[{\citenamefont{Appelquist et~al.}(1988)\citenamefont{Appelquist, Nash,
  and Wijewardhana}}]{Appelquist:1988sr}
\bibinfo{author}{\bibfnamefont{T.}~\bibnamefont{Appelquist}},
  \bibinfo{author}{\bibfnamefont{D.}~\bibnamefont{Nash}}, \bibnamefont{and}
  \bibinfo{author}{\bibfnamefont{L.~C.~R.} \bibnamefont{Wijewardhana}},
  \bibinfo{journal}{Phys. Rev. Lett.} \textbf{\bibinfo{volume}{60}},
  \bibinfo{pages}{2575} (\bibinfo{year}{1988}).

\bibitem[{\citenamefont{Atkinson et~al.}(1990)\citenamefont{Atkinson, Johnson,
  and Maris}}]{Atkinson:1989fp}
\bibinfo{author}{\bibfnamefont{D.}~\bibnamefont{Atkinson}},
  \bibinfo{author}{\bibfnamefont{P.~W.} \bibnamefont{Johnson}},
  \bibnamefont{and} \bibinfo{author}{\bibfnamefont{P.}~\bibnamefont{Maris}},
  \bibinfo{journal}{Phys. Rev.} \textbf{\bibinfo{volume}{D42}},
  \bibinfo{pages}{602} (\bibinfo{year}{1990}).

\bibitem[{\citenamefont{Pennington and Walsh}(1991)}]{Pennington:1990bx}
\bibinfo{author}{\bibfnamefont{M.~R.} \bibnamefont{Pennington}}
  \bibnamefont{and} \bibinfo{author}{\bibfnamefont{D.}~\bibnamefont{Walsh}},
  \bibinfo{journal}{Phys. Lett.} \textbf{\bibinfo{volume}{B253}},
  \bibinfo{pages}{246} (\bibinfo{year}{1991}).

\bibitem[{\citenamefont{Curtis et~al.}(1992)\citenamefont{Curtis, Pennington,
  and Walsh}}]{Curtis:1992gm}
\bibinfo{author}{\bibfnamefont{D.~C.} \bibnamefont{Curtis}},
  \bibinfo{author}{\bibfnamefont{M.~R.} \bibnamefont{Pennington}},
  \bibnamefont{and} \bibinfo{author}{\bibfnamefont{D.}~\bibnamefont{Walsh}},
  \bibinfo{journal}{Phys. Lett.} \textbf{\bibinfo{volume}{B295}},
  \bibinfo{pages}{313} (\bibinfo{year}{1992}).

\bibitem[{\citenamefont{Burden and Roberts}(1991)}]{Burden:1990mg}
\bibinfo{author}{\bibfnamefont{C.~J.} \bibnamefont{Burden}} \bibnamefont{and}
  \bibinfo{author}{\bibfnamefont{C.~D.} \bibnamefont{Roberts}},
  \bibinfo{journal}{Phys. Rev.} \textbf{\bibinfo{volume}{D44}},
  \bibinfo{pages}{540} (\bibinfo{year}{1991}).

\bibitem[{\citenamefont{Maris}(1995)}]{Maris:1995ns}
\bibinfo{author}{\bibfnamefont{P.}~\bibnamefont{Maris}},
  \bibinfo{journal}{Phys. Rev.} \textbf{\bibinfo{volume}{D52}},
  \bibinfo{pages}{6087} (\bibinfo{year}{1995}), \eprint{hep-ph/9508323}.

\bibitem[{\citenamefont{Gusynin et~al.}(1996)\citenamefont{Gusynin, Hams, and
  Reenders}}]{Gusynin:1995bb}
\bibinfo{author}{\bibfnamefont{V.~P.} \bibnamefont{Gusynin}},
  \bibinfo{author}{\bibfnamefont{A.~H.} \bibnamefont{Hams}}, \bibnamefont{and}
  \bibinfo{author}{\bibfnamefont{M.}~\bibnamefont{Reenders}},
  \bibinfo{journal}{Phys. Rev.} \textbf{\bibinfo{volume}{D53}},
  \bibinfo{pages}{2227} (\bibinfo{year}{1996}), \eprint{hep-ph/9509380}.

\bibitem[{\citenamefont{Maris}(1996)}]{Maris:1996zg}
\bibinfo{author}{\bibfnamefont{P.}~\bibnamefont{Maris}},
  \bibinfo{journal}{Phys. Rev.} \textbf{\bibinfo{volume}{D54}},
  \bibinfo{pages}{4049} (\bibinfo{year}{1996}), \eprint{hep-ph/9606214}.

\bibitem[{\citenamefont{Fischer et~al.}(2004)\citenamefont{Fischer, Alkofer,
  Dahm, and Maris}}]{Fischer:2004nq}
\bibinfo{author}{\bibfnamefont{C.~S.} \bibnamefont{Fischer}},
  \bibinfo{author}{\bibfnamefont{R.}~\bibnamefont{Alkofer}},
  \bibinfo{author}{\bibfnamefont{T.}~\bibnamefont{Dahm}}, \bibnamefont{and}
  \bibinfo{author}{\bibfnamefont{P.}~\bibnamefont{Maris}},
  \bibinfo{journal}{Phys. Rev.} \textbf{\bibinfo{volume}{D70}},
  \bibinfo{pages}{073007} (\bibinfo{year}{2004}), \eprint{hep-ph/0407104}.

\bibitem[{\citenamefont{Dagotto et~al.}(1990)\citenamefont{Dagotto, Kocic, and
  Kogut}}]{Dagotto:1989td}
\bibinfo{author}{\bibfnamefont{E.}~\bibnamefont{Dagotto}},
  \bibinfo{author}{\bibfnamefont{A.}~\bibnamefont{Kocic}}, \bibnamefont{and}
  \bibinfo{author}{\bibfnamefont{J.~B.} \bibnamefont{Kogut}},
  \bibinfo{journal}{Nucl. Phys.} \textbf{\bibinfo{volume}{B334}},
  \bibinfo{pages}{279} (\bibinfo{year}{1990}).

\bibitem[{\citenamefont{Hands and Kogut}(1990)}]{Hands:1989mv}
\bibinfo{author}{\bibfnamefont{S.}~\bibnamefont{Hands}} \bibnamefont{and}
  \bibinfo{author}{\bibfnamefont{J.~B.} \bibnamefont{Kogut}},
  \bibinfo{journal}{Nucl. Phys.} \textbf{\bibinfo{volume}{B335}},
  \bibinfo{pages}{455} (\bibinfo{year}{1990}).

\bibitem[{\citenamefont{Hands et~al.}(2002)\citenamefont{Hands, Kogut, and
  Strouthos}}]{Hands:2002dv}
\bibinfo{author}{\bibfnamefont{S.~J.} \bibnamefont{Hands}},
  \bibinfo{author}{\bibfnamefont{J.~B.} \bibnamefont{Kogut}}, \bibnamefont{and}
  \bibinfo{author}{\bibfnamefont{C.~G.} \bibnamefont{Strouthos}},
  \bibinfo{journal}{Nucl. Phys.} \textbf{\bibinfo{volume}{B645}},
  \bibinfo{pages}{321} (\bibinfo{year}{2002}), \eprint{hep-lat/0208030}.

\bibitem[{\citenamefont{Hands et~al.}(2004)\citenamefont{Hands, Kogut,
  Scorzato, and Strouthos}}]{Hands:2004bh}
\bibinfo{author}{\bibfnamefont{S.~J.} \bibnamefont{Hands}},
  \bibinfo{author}{\bibfnamefont{J.~B.} \bibnamefont{Kogut}},
  \bibinfo{author}{\bibfnamefont{L.}~\bibnamefont{Scorzato}}, \bibnamefont{and}
  \bibinfo{author}{\bibfnamefont{C.~G.} \bibnamefont{Strouthos}},
  \bibinfo{journal}{Phys. Rev.} \textbf{\bibinfo{volume}{B70}},
  \bibinfo{pages}{104501} (\bibinfo{year}{2004}), \eprint{hep-lat/0404013}.

\bibitem[{\citenamefont{Miransky and Yamawaki}(1997)}]{Miransky:1996pd}
\bibinfo{author}{\bibfnamefont{V.~A.} \bibnamefont{Miransky}} \bibnamefont{and}
  \bibinfo{author}{\bibfnamefont{K.}~\bibnamefont{Yamawaki}},
  \bibinfo{journal}{Phys. Rev.} \textbf{\bibinfo{volume}{D55}},
  \bibinfo{pages}{5051} (\bibinfo{year}{1997}), \eprint{hep-th/9611142}.

\bibitem[{\citenamefont{Appelquist et~al.}(1996)\citenamefont{Appelquist,
  Terning, and Wijewardhana}}]{Appelquist:1996dq}
\bibinfo{author}{\bibfnamefont{T.}~\bibnamefont{Appelquist}},
  \bibinfo{author}{\bibfnamefont{J.}~\bibnamefont{Terning}}, \bibnamefont{and}
  \bibinfo{author}{\bibfnamefont{L.~C.~R.} \bibnamefont{Wijewardhana}},
  \bibinfo{journal}{Phys. Rev. Lett.} \textbf{\bibinfo{volume}{77}},
  \bibinfo{pages}{1214} (\bibinfo{year}{1996}), \eprint{hep-ph/9602385}.

\bibitem[{\citenamefont{Appelquist and Selipsky}(1997)}]{Appelquist:1997dc}
\bibinfo{author}{\bibfnamefont{T.}~\bibnamefont{Appelquist}} \bibnamefont{and}
  \bibinfo{author}{\bibfnamefont{S.~B.} \bibnamefont{Selipsky}},
  \bibinfo{journal}{Phys. Lett.} \textbf{\bibinfo{volume}{B400}},
  \bibinfo{pages}{364} (\bibinfo{year}{1997}), \eprint{hep-ph/9702404}.

\bibitem[{\citenamefont{Schafer and Shuryak}(1998)}]{Schafer:1996wv}
\bibinfo{author}{\bibfnamefont{T.}~\bibnamefont{Schafer}} \bibnamefont{and}
  \bibinfo{author}{\bibfnamefont{E.~V.} \bibnamefont{Shuryak}},
  \bibinfo{journal}{Rev. Mod. Phys.} \textbf{\bibinfo{volume}{70}},
  \bibinfo{pages}{323} (\bibinfo{year}{1998}), \eprint{hep-ph/9610451}.

\bibitem[{\citenamefont{Velkovsky and Shuryak}(1998)}]{Velkovsky:1997fe}
\bibinfo{author}{\bibfnamefont{M.}~\bibnamefont{Velkovsky}} \bibnamefont{and}
  \bibinfo{author}{\bibfnamefont{E.~V.} \bibnamefont{Shuryak}},
  \bibinfo{journal}{Phys. Lett.} \textbf{\bibinfo{volume}{B437}},
  \bibinfo{pages}{398} (\bibinfo{year}{1998}), \eprint{hep-ph/9703345}.

\bibitem[{\citenamefont{Appelquist et~al.}(1998)\citenamefont{Appelquist,
  Ratnaweera, Terning, and Wijewardhana}}]{Appelquist:1998rb}
\bibinfo{author}{\bibfnamefont{T.}~\bibnamefont{Appelquist}},
  \bibinfo{author}{\bibfnamefont{A.}~\bibnamefont{Ratnaweera}},
  \bibinfo{author}{\bibfnamefont{J.}~\bibnamefont{Terning}}, \bibnamefont{and}
  \bibinfo{author}{\bibfnamefont{L.~C.~R.} \bibnamefont{Wijewardhana}},
  \bibinfo{journal}{Phys. Rev.} \textbf{\bibinfo{volume}{D58}},
  \bibinfo{pages}{105017} (\bibinfo{year}{1998}), \eprint{hep-ph/9806472}.

\bibitem[{\citenamefont{Harada and Yamawaki}(2001)}]{Harada:2000kb}
\bibinfo{author}{\bibfnamefont{M.}~\bibnamefont{Harada}} \bibnamefont{and}
  \bibinfo{author}{\bibfnamefont{K.}~\bibnamefont{Yamawaki}},
  \bibinfo{journal}{Phys. Rev. Lett.} \textbf{\bibinfo{volume}{86}},
  \bibinfo{pages}{757} (\bibinfo{year}{2001}), \eprint{hep-ph/0010207}.

\bibitem[{\citenamefont{Sannino and Schechter}(1999)}]{Sannino:1999qe}
\bibinfo{author}{\bibfnamefont{F.}~\bibnamefont{Sannino}} \bibnamefont{and}
  \bibinfo{author}{\bibfnamefont{J.}~\bibnamefont{Schechter}},
  \bibinfo{journal}{Phys. Rev.} \textbf{\bibinfo{volume}{D60}},
  \bibinfo{pages}{056004} (\bibinfo{year}{1999}), \eprint{hep-ph/9903359}.

\bibitem[{\citenamefont{Harada et~al.}(2003)\citenamefont{Harada, Kurachi, and
  Yamawaki}}]{Harada:2003dc}
\bibinfo{author}{\bibfnamefont{M.}~\bibnamefont{Harada}},
  \bibinfo{author}{\bibfnamefont{M.}~\bibnamefont{Kurachi}}, \bibnamefont{and}
  \bibinfo{author}{\bibfnamefont{K.}~\bibnamefont{Yamawaki}},
  \bibinfo{journal}{Phys. Rev.} \textbf{\bibinfo{volume}{D68}},
  \bibinfo{pages}{076001} (\bibinfo{year}{2003}), \eprint{hep-ph/0305018}.

\bibitem[{\citenamefont{Gies and Jaeckel}(2006)}]{Gies:2005as}
\bibinfo{author}{\bibfnamefont{H.}~\bibnamefont{Gies}} \bibnamefont{and}
  \bibinfo{author}{\bibfnamefont{J.}~\bibnamefont{Jaeckel}},
  \bibinfo{journal}{Eur. Phys. J.} \textbf{\bibinfo{volume}{C46}},
  \bibinfo{pages}{433} (\bibinfo{year}{2006}), \eprint{hep-ph/0507171}.

\bibitem[{\citenamefont{Braun and Gies}(2007)}]{Braun:2005uj}
\bibinfo{author}{\bibfnamefont{J.}~\bibnamefont{Braun}} \bibnamefont{and}
  \bibinfo{author}{\bibfnamefont{H.}~\bibnamefont{Gies}},
  \bibinfo{journal}{Phys. Lett.} \textbf{\bibinfo{volume}{B645}},
  \bibinfo{pages}{53} (\bibinfo{year}{2007}), \eprint{hep-ph/0512085}.

\bibitem[{\citenamefont{Braun and Gies}(2006)}]{Braun:2006jd}
\bibinfo{author}{\bibfnamefont{J.}~\bibnamefont{Braun}} \bibnamefont{and}
  \bibinfo{author}{\bibfnamefont{H.}~\bibnamefont{Gies}},
  \bibinfo{journal}{JHEP} \textbf{\bibinfo{volume}{06}}, \bibinfo{pages}{024}
  (\bibinfo{year}{2006}), \eprint{hep-ph/0602226}.

\bibitem[{\citenamefont{Terao and Tsuchiya}(2007)}]{Terao:2007jm}
\bibinfo{author}{\bibfnamefont{H.}~\bibnamefont{Terao}} \bibnamefont{and}
  \bibinfo{author}{\bibfnamefont{A.}~\bibnamefont{Tsuchiya}}
  (\bibinfo{year}{2007}), \eprint{0704.3659}.

\bibitem[{\citenamefont{Poppitz and Unsal}(2009)}]{Poppitz:2009uq}
\bibinfo{author}{\bibfnamefont{E.}~\bibnamefont{Poppitz}} \bibnamefont{and}
  \bibinfo{author}{\bibfnamefont{M.}~\bibnamefont{Unsal}},
  \bibinfo{journal}{JHEP} \textbf{\bibinfo{volume}{09}}, \bibinfo{pages}{050}
  (\bibinfo{year}{2009}), \eprint{0906.5156}.

\bibitem[{\citenamefont{Armoni}(2010)}]{Armoni:2009jn}
\bibinfo{author}{\bibfnamefont{A.}~\bibnamefont{Armoni}},
  \bibinfo{journal}{Nucl. Phys.} \textbf{\bibinfo{volume}{B826}},
  \bibinfo{pages}{328} (\bibinfo{year}{2010}), \eprint{0907.4091}.

\bibitem[{\citenamefont{Sannino}(2009{\natexlab{b}})}]{Sannino:2009qc}
\bibinfo{author}{\bibfnamefont{F.}~\bibnamefont{Sannino}},
  \bibinfo{journal}{Phys. Rev.} \textbf{\bibinfo{volume}{D80}},
  \bibinfo{pages}{065011} (\bibinfo{year}{2009}{\natexlab{b}}),
  \eprint{0907.1364}.

\bibitem[{\citenamefont{Sannino}(2010)}]{Sannino:2009me}
\bibinfo{author}{\bibfnamefont{F.}~\bibnamefont{Sannino}},
  \bibinfo{journal}{Nucl. Phys.} \textbf{\bibinfo{volume}{B830}},
  \bibinfo{pages}{179} (\bibinfo{year}{2010}), \eprint{0909.4584}.

\bibitem[{\citenamefont{Kogut et~al.}(1982)}]{Kogut:1982fn}
\bibinfo{author}{\bibfnamefont{J.~B.} \bibnamefont{Kogut}}
  \bibnamefont{et~al.}, \bibinfo{journal}{Phys. Rev. Lett.}
  \textbf{\bibinfo{volume}{48}}, \bibinfo{pages}{1140} (\bibinfo{year}{1982}).

\bibitem[{\citenamefont{Gavai}(1986)}]{Gavai:1985wi}
\bibinfo{author}{\bibfnamefont{R.~V.} \bibnamefont{Gavai}},
  \bibinfo{journal}{Nucl. Phys.} \textbf{\bibinfo{volume}{B269}},
  \bibinfo{pages}{530} (\bibinfo{year}{1986}).

\bibitem[{\citenamefont{Fukugita et~al.}(1988)\citenamefont{Fukugita, Ohta, and
  Ukawa}}]{Fukugita:1987mb}
\bibinfo{author}{\bibfnamefont{M.}~\bibnamefont{Fukugita}},
  \bibinfo{author}{\bibfnamefont{S.}~\bibnamefont{Ohta}}, \bibnamefont{and}
  \bibinfo{author}{\bibfnamefont{A.}~\bibnamefont{Ukawa}},
  \bibinfo{journal}{Phys. Rev. Lett.} \textbf{\bibinfo{volume}{60}},
  \bibinfo{pages}{178} (\bibinfo{year}{1988}).

\bibitem[{\citenamefont{Brown et~al.}(1992)}]{Brown:1992fz}
\bibinfo{author}{\bibfnamefont{F.~R.} \bibnamefont{Brown}}
  \bibnamefont{et~al.}, \bibinfo{journal}{Phys. Rev.}
  \textbf{\bibinfo{volume}{D46}}, \bibinfo{pages}{5655} (\bibinfo{year}{1992}),
  \eprint{hep-lat/9206001}.

\bibitem[{\citenamefont{Damgaard et~al.}(1997)\citenamefont{Damgaard, Heller,
  Krasnitz, and Olesen}}]{Damgaard:1997ut}
\bibinfo{author}{\bibfnamefont{P.~H.} \bibnamefont{Damgaard}},
  \bibinfo{author}{\bibfnamefont{U.~M.} \bibnamefont{Heller}},
  \bibinfo{author}{\bibfnamefont{A.}~\bibnamefont{Krasnitz}}, \bibnamefont{and}
  \bibinfo{author}{\bibfnamefont{P.}~\bibnamefont{Olesen}},
  \bibinfo{journal}{Phys. Lett.} \textbf{\bibinfo{volume}{B400}},
  \bibinfo{pages}{169} (\bibinfo{year}{1997}), \eprint{hep-lat/9701008}.

\bibitem[{\citenamefont{Iwasaki et~al.}(2004)\citenamefont{Iwasaki, Kanaya,
  Kaya, Sakai, and Yoshie}}]{Iwasaki:2003de}
\bibinfo{author}{\bibfnamefont{Y.}~\bibnamefont{Iwasaki}},
  \bibinfo{author}{\bibfnamefont{K.}~\bibnamefont{Kanaya}},
  \bibinfo{author}{\bibfnamefont{S.}~\bibnamefont{Kaya}},
  \bibinfo{author}{\bibfnamefont{S.}~\bibnamefont{Sakai}}, \bibnamefont{and}
  \bibinfo{author}{\bibfnamefont{T.}~\bibnamefont{Yoshie}},
  \bibinfo{journal}{Phys. Rev.} \textbf{\bibinfo{volume}{D69}},
  \bibinfo{pages}{014507} (\bibinfo{year}{2004}), \eprint{hep-lat/0309159}.

\bibitem[{\citenamefont{Catterall and Sannino}(2007)}]{Catterall:2007yx}
\bibinfo{author}{\bibfnamefont{S.}~\bibnamefont{Catterall}} \bibnamefont{and}
  \bibinfo{author}{\bibfnamefont{F.}~\bibnamefont{Sannino}},
  \bibinfo{journal}{Phys. Rev.} \textbf{\bibinfo{volume}{D76}},
  \bibinfo{pages}{034504} (\bibinfo{year}{2007}), \eprint{0705.1664}.

\bibitem[{\citenamefont{Appelquist et~al.}(2008)\citenamefont{Appelquist,
  Fleming, and Neil}}]{Appelquist:2007hu}
\bibinfo{author}{\bibfnamefont{T.}~\bibnamefont{Appelquist}},
  \bibinfo{author}{\bibfnamefont{G.~T.} \bibnamefont{Fleming}},
  \bibnamefont{and} \bibinfo{author}{\bibfnamefont{E.~T.} \bibnamefont{Neil}},
  \bibinfo{journal}{Phys. Rev. Lett.} \textbf{\bibinfo{volume}{100}},
  \bibinfo{pages}{171607} (\bibinfo{year}{2008}), \eprint{0712.0609}.

\bibitem[{\citenamefont{Deuzeman et~al.}(2008)\citenamefont{Deuzeman, Lombardo,
  and Pallante}}]{Deuzeman:2008sc}
\bibinfo{author}{\bibfnamefont{A.}~\bibnamefont{Deuzeman}},
  \bibinfo{author}{\bibfnamefont{M.~P.} \bibnamefont{Lombardo}},
  \bibnamefont{and} \bibinfo{author}{\bibfnamefont{E.}~\bibnamefont{Pallante}},
  \bibinfo{journal}{Phys. Lett.} \textbf{\bibinfo{volume}{B670}},
  \bibinfo{pages}{41} (\bibinfo{year}{2008}), \eprint{0804.2905}.

\bibitem[{\citenamefont{Deuzeman et~al.}(2009)\citenamefont{Deuzeman, Lombardo,
  and Pallante}}]{Deuzeman:2009mh}
\bibinfo{author}{\bibfnamefont{A.}~\bibnamefont{Deuzeman}},
  \bibinfo{author}{\bibfnamefont{M.~P.} \bibnamefont{Lombardo}},
  \bibnamefont{and} \bibinfo{author}{\bibfnamefont{E.}~\bibnamefont{Pallante}}
  (\bibinfo{year}{2009}), \eprint{0904.4662}.

\bibitem[{\citenamefont{Appelquist et~al.}(2009)\citenamefont{Appelquist,
  Fleming, and Neil}}]{Appelquist:2009ty}
\bibinfo{author}{\bibfnamefont{T.}~\bibnamefont{Appelquist}},
  \bibinfo{author}{\bibfnamefont{G.~T.} \bibnamefont{Fleming}},
  \bibnamefont{and} \bibinfo{author}{\bibfnamefont{E.~T.} \bibnamefont{Neil}},
  \bibinfo{journal}{Phys. Rev.} \textbf{\bibinfo{volume}{D79}},
  \bibinfo{pages}{076010} (\bibinfo{year}{2009}), \eprint{0901.3766}.

\bibitem[{\citenamefont{Fodor et~al.}(2009{\natexlab{a}})\citenamefont{Fodor,
  Holland, Kuti, Nogradi, and Schroeder}}]{Fodor:2009wk}
\bibinfo{author}{\bibfnamefont{Z.}~\bibnamefont{Fodor}},
  \bibinfo{author}{\bibfnamefont{K.}~\bibnamefont{Holland}},
  \bibinfo{author}{\bibfnamefont{J.}~\bibnamefont{Kuti}},
  \bibinfo{author}{\bibfnamefont{D.}~\bibnamefont{Nogradi}}, \bibnamefont{and}
  \bibinfo{author}{\bibfnamefont{C.}~\bibnamefont{Schroeder}},
  \bibinfo{journal}{Phys. Lett.} \textbf{\bibinfo{volume}{B681}},
  \bibinfo{pages}{353} (\bibinfo{year}{2009}{\natexlab{a}}),
  \eprint{0907.4562}.

\bibitem[{\citenamefont{Fodor et~al.}(2009{\natexlab{b}})\citenamefont{Fodor,
  Holland, Kuti, Nogradi, and Schroeder}}]{Fodor:2009ff}
\bibinfo{author}{\bibfnamefont{Z.}~\bibnamefont{Fodor}},
  \bibinfo{author}{\bibfnamefont{K.}~\bibnamefont{Holland}},
  \bibinfo{author}{\bibfnamefont{J.}~\bibnamefont{Kuti}},
  \bibinfo{author}{\bibfnamefont{D.}~\bibnamefont{Nogradi}}, \bibnamefont{and}
  \bibinfo{author}{\bibfnamefont{C.}~\bibnamefont{Schroeder}}
  (\bibinfo{year}{2009}{\natexlab{b}}), \eprint{0911.2463}.

\bibitem[{\citenamefont{Pallante}(2009)}]{Pallante:2009hu}
\bibinfo{author}{\bibfnamefont{E.}~\bibnamefont{Pallante}}
  (\bibinfo{year}{2009}), \eprint{0912.5188}.

\bibitem[{\citenamefont{DeGrand}(2010)}]{DeGrand:2010ba}
\bibinfo{author}{\bibfnamefont{T.}~\bibnamefont{DeGrand}}
  (\bibinfo{year}{2010}), \eprint{1010.4741}.

\bibitem[{\citenamefont{Miransky and Yamawaki}(1989)}]{Miransky:1988gk}
\bibinfo{author}{\bibfnamefont{V.~A.} \bibnamefont{Miransky}} \bibnamefont{and}
  \bibinfo{author}{\bibfnamefont{K.}~\bibnamefont{Yamawaki}},
  \bibinfo{journal}{Mod. Phys. Lett.} \textbf{\bibinfo{volume}{A4}},
  \bibinfo{pages}{129} (\bibinfo{year}{1989}).

\bibitem[{\citenamefont{Chivukula}(1997)}]{Chivukula:1996kg}
\bibinfo{author}{\bibfnamefont{R.~S.} \bibnamefont{Chivukula}},
  \bibinfo{journal}{Phys. Rev.} \textbf{\bibinfo{volume}{D55}},
  \bibinfo{pages}{5238} (\bibinfo{year}{1997}), \eprint{hep-ph/9612267}.

\bibitem[{\citenamefont{Appelquist and Sannino}(1999)}]{Appelquist:1998xf}
\bibinfo{author}{\bibfnamefont{T.}~\bibnamefont{Appelquist}} \bibnamefont{and}
  \bibinfo{author}{\bibfnamefont{F.}~\bibnamefont{Sannino}},
  \bibinfo{journal}{Phys. Rev.} \textbf{\bibinfo{volume}{D59}},
  \bibinfo{pages}{067702} (\bibinfo{year}{1999}), \eprint{hep-ph/9806409}.

\bibitem[{\citenamefont{Berezinskii}(1971)}]{Berezinskii}
\bibinfo{author}{\bibfnamefont{V.~L.} \bibnamefont{Berezinskii}},
  \bibinfo{journal}{Sov. Phys. JETP} \textbf{\bibinfo{volume}{32}},
  \bibinfo{pages}{493} (\bibinfo{year}{1971}).

\bibitem[{\citenamefont{Berezinskii}(1972)}]{Berezinskii2}
\bibinfo{author}{\bibfnamefont{V.~L.} \bibnamefont{Berezinskii}},
  \bibinfo{journal}{Sov. Phys. JETP} \textbf{\bibinfo{volume}{34}},
  \bibinfo{pages}{610} (\bibinfo{year}{1972}).

\bibitem[{\citenamefont{Kosterlitz and Thouless}(1973)}]{Kosterlitz:1973xp}
\bibinfo{author}{\bibfnamefont{J.~M.} \bibnamefont{Kosterlitz}}
  \bibnamefont{and} \bibinfo{author}{\bibfnamefont{D.~J.}
  \bibnamefont{Thouless}}, \bibinfo{journal}{J. Phys.}
  \textbf{\bibinfo{volume}{C6}}, \bibinfo{pages}{1181} (\bibinfo{year}{1973}).

\bibitem[{\citenamefont{Kaplan et~al.}(2009)\citenamefont{Kaplan, Lee, Son, and
  Stephanov}}]{Kaplan:2009kr}
\bibinfo{author}{\bibfnamefont{D.~B.} \bibnamefont{Kaplan}},
  \bibinfo{author}{\bibfnamefont{J.-W.} \bibnamefont{Lee}},
  \bibinfo{author}{\bibfnamefont{D.~T.} \bibnamefont{Son}}, \bibnamefont{and}
  \bibinfo{author}{\bibfnamefont{M.~A.} \bibnamefont{Stephanov}},
  \bibinfo{journal}{Phys. Rev.} \textbf{\bibinfo{volume}{D80}},
  \bibinfo{pages}{125005} (\bibinfo{year}{2009}), \eprint{0905.4752}.

\bibitem[{\citenamefont{Braun and Gies}(2010)}]{Braun:2009ns}
\bibinfo{author}{\bibfnamefont{J.}~\bibnamefont{Braun}} \bibnamefont{and}
  \bibinfo{author}{\bibfnamefont{H.}~\bibnamefont{Gies}},
  \bibinfo{journal}{JHEP} \textbf{\bibinfo{volume}{05}}, \bibinfo{pages}{060}
  (\bibinfo{year}{2010}), \eprint{0912.4168}.

\bibitem[{\citenamefont{Jarvinen and Sannino}(2010)}]{Jarvinen:2010ks}
\bibinfo{author}{\bibfnamefont{M.}~\bibnamefont{Jarvinen}} \bibnamefont{and}
  \bibinfo{author}{\bibfnamefont{F.}~\bibnamefont{Sannino}}
  (\bibinfo{year}{2010}), \eprint{1009.5380}.

\bibitem[{\citenamefont{DeGrand and Hasenfratz}(2009)}]{DeGrand:2009mt}
\bibinfo{author}{\bibfnamefont{T.}~\bibnamefont{DeGrand}} \bibnamefont{and}
  \bibinfo{author}{\bibfnamefont{A.}~\bibnamefont{Hasenfratz}},
  \bibinfo{journal}{Phys. Rev.} \textbf{\bibinfo{volume}{D80}},
  \bibinfo{pages}{034506} (\bibinfo{year}{2009}), \eprint{0906.1976}.

\bibitem[{\citenamefont{Del~Debbio and
  Zwicky}(2010{\natexlab{a}})}]{DelDebbio:2010ze}
\bibinfo{author}{\bibfnamefont{L.}~\bibnamefont{Del~Debbio}} \bibnamefont{and}
  \bibinfo{author}{\bibfnamefont{R.}~\bibnamefont{Zwicky}},
  \bibinfo{journal}{Phys. Rev.} \textbf{\bibinfo{volume}{D82}},
  \bibinfo{pages}{014502} (\bibinfo{year}{2010}{\natexlab{a}}),
  \eprint{1005.2371}.

\bibitem[{\citenamefont{Del~Debbio and
  Zwicky}(2010{\natexlab{b}})}]{DelDebbio:2010jy}
\bibinfo{author}{\bibfnamefont{L.}~\bibnamefont{Del~Debbio}} \bibnamefont{and}
  \bibinfo{author}{\bibfnamefont{R.}~\bibnamefont{Zwicky}}
  (\bibinfo{year}{2010}{\natexlab{b}}), \eprint{1009.2894}.

\bibitem[{\citenamefont{Braun}(2006)}]{Braun:2006wu}
\bibinfo{author}{\bibfnamefont{J.}~\bibnamefont{Braun}} (\bibinfo{year}{2006}),
  \eprint{hep-ph/0611145}.

\bibitem[{\citenamefont{Gies et~al.}(2004)\citenamefont{Gies, Jaeckel, and
  Wetterich}}]{Gies:2003dp}
\bibinfo{author}{\bibfnamefont{H.}~\bibnamefont{Gies}},
  \bibinfo{author}{\bibfnamefont{J.}~\bibnamefont{Jaeckel}}, \bibnamefont{and}
  \bibinfo{author}{\bibfnamefont{C.}~\bibnamefont{Wetterich}},
  \bibinfo{journal}{Phys. Rev.} \textbf{\bibinfo{volume}{D69}},
  \bibinfo{pages}{105008} (\bibinfo{year}{2004}), \eprint{hep-ph/0312034}.

\bibitem[{\citenamefont{Fukano and Sannino}(2010)}]{Fukano:2010yv}
\bibinfo{author}{\bibfnamefont{H.~S.} \bibnamefont{Fukano}} \bibnamefont{and}
  \bibinfo{author}{\bibfnamefont{F.}~\bibnamefont{Sannino}},
  \bibinfo{journal}{Phys. Rev.} \textbf{\bibinfo{volume}{D82}},
  \bibinfo{pages}{035021} (\bibinfo{year}{2010}), \eprint{1005.3340}.

\bibitem[{\citenamefont{Braun et~al.}(2010)\citenamefont{Braun, Gies, and
  Scherer}}]{Braun:2010tt}
\bibinfo{author}{\bibfnamefont{J.}~\bibnamefont{Braun}},
  \bibinfo{author}{\bibfnamefont{H.}~\bibnamefont{Gies}}, \bibnamefont{and}
  \bibinfo{author}{\bibfnamefont{D.~D.} \bibnamefont{Scherer}}
  (\bibinfo{year}{2010}), \eprint{1011.1456}.

\bibitem[{\citenamefont{Braun}(2010)}]{Braun:2009si}
\bibinfo{author}{\bibfnamefont{J.}~\bibnamefont{Braun}},
  \bibinfo{journal}{Phys. Rev.} \textbf{\bibinfo{volume}{D81}},
  \bibinfo{pages}{016008} (\bibinfo{year}{2010}), \eprint{0908.1543}.

\bibitem[{\citenamefont{Pisarski and Wilczek}(1984)}]{Pisarski:1983ms}
\bibinfo{author}{\bibfnamefont{R.~D.} \bibnamefont{Pisarski}} \bibnamefont{and}
  \bibinfo{author}{\bibfnamefont{F.}~\bibnamefont{Wilczek}},
  \bibinfo{journal}{Phys. Rev.} \textbf{\bibinfo{volume}{D29}},
  \bibinfo{pages}{338} (\bibinfo{year}{1984}).

\bibitem[{\citenamefont{Wetterich}(1993)}]{Wetterich:1992yh}
\bibinfo{author}{\bibfnamefont{C.}~\bibnamefont{Wetterich}},
  \bibinfo{journal}{Phys. Lett.} \textbf{\bibinfo{volume}{B301}},
  \bibinfo{pages}{90} (\bibinfo{year}{1993}).

\bibitem[{\citenamefont{Reuter}(1996)}]{Reuter:1996ub}
\bibinfo{author}{\bibfnamefont{M.}~\bibnamefont{Reuter}}
  (\bibinfo{year}{1996}), \eprint{hep-th/9602012}.

\bibitem[{\citenamefont{Litim and Pawlowski}(1998)}]{Litim:1998nf}
\bibinfo{author}{\bibfnamefont{D.~F.} \bibnamefont{Litim}} \bibnamefont{and}
  \bibinfo{author}{\bibfnamefont{J.~M.} \bibnamefont{Pawlowski}}
  (\bibinfo{year}{1998}), \eprint{hep-th/9901063}.

\bibitem[{\citenamefont{Bagnuls and Bervillier}(2001)}]{Bagnuls:2000ae}
\bibinfo{author}{\bibfnamefont{C.}~\bibnamefont{Bagnuls}} \bibnamefont{and}
  \bibinfo{author}{\bibfnamefont{C.}~\bibnamefont{Bervillier}},
  \bibinfo{journal}{Phys. Rept.} \textbf{\bibinfo{volume}{348}},
  \bibinfo{pages}{91} (\bibinfo{year}{2001}), \eprint{hep-th/0002034}.

\bibitem[{\citenamefont{Berges et~al.}(2002)\citenamefont{Berges, Tetradis, and
  Wetterich}}]{Berges:2000ew}
\bibinfo{author}{\bibfnamefont{J.}~\bibnamefont{Berges}},
  \bibinfo{author}{\bibfnamefont{N.}~\bibnamefont{Tetradis}}, \bibnamefont{and}
  \bibinfo{author}{\bibfnamefont{C.}~\bibnamefont{Wetterich}},
  \bibinfo{journal}{Phys. Rept.} \textbf{\bibinfo{volume}{363}},
  \bibinfo{pages}{223} (\bibinfo{year}{2002}), \eprint{hep-ph/0005122}.

\bibitem[{\citenamefont{Polonyi}(2003)}]{Polonyi:2001se}
\bibinfo{author}{\bibfnamefont{J.}~\bibnamefont{Polonyi}},
  \bibinfo{journal}{Central Eur. J. Phys.} \textbf{\bibinfo{volume}{1}},
  \bibinfo{pages}{1} (\bibinfo{year}{2003}), \eprint{hep-th/0110026}.

\bibitem[{\citenamefont{Delamotte et~al.}(2004)\citenamefont{Delamotte,
  Mouhanna, and Tissier}}]{Delamotte:2003dw}
\bibinfo{author}{\bibfnamefont{B.}~\bibnamefont{Delamotte}},
  \bibinfo{author}{\bibfnamefont{D.}~\bibnamefont{Mouhanna}}, \bibnamefont{and}
  \bibinfo{author}{\bibfnamefont{M.}~\bibnamefont{Tissier}},
  \bibinfo{journal}{Phys. Rev.} \textbf{\bibinfo{volume}{B69}},
  \bibinfo{pages}{134413} (\bibinfo{year}{2004}), \eprint{cond-mat/0309101}.

\bibitem[{\citenamefont{Pawlowski}(2007)}]{Pawlowski:2005xe}
\bibinfo{author}{\bibfnamefont{J.~M.} \bibnamefont{Pawlowski}},
  \bibinfo{journal}{Annals Phys.} \textbf{\bibinfo{volume}{322}},
  \bibinfo{pages}{2831} (\bibinfo{year}{2007}), \eprint{hep-th/0512261}.

\bibitem[{\citenamefont{Gies}(2006)}]{Gies:2006wv}
\bibinfo{author}{\bibfnamefont{H.}~\bibnamefont{Gies}} (\bibinfo{year}{2006}),
  \eprint{hep-ph/0611146}.

\bibitem[{\citenamefont{Delamotte}(2007)}]{Delamotte:2007pf}
\bibinfo{author}{\bibfnamefont{B.}~\bibnamefont{Delamotte}}
  (\bibinfo{year}{2007}), \eprint{cond-mat/0702365}.

\bibitem[{\citenamefont{Rosten}(2010)}]{Rosten:2010vm}
\bibinfo{author}{\bibfnamefont{O.~J.} \bibnamefont{Rosten}}
  (\bibinfo{year}{2010}), \eprint{1003.1366}.

\bibitem[{\citenamefont{Jungnickel and Wetterich}(1996)}]{Jungnickel:1995fp}
\bibinfo{author}{\bibfnamefont{D.~U.} \bibnamefont{Jungnickel}}
  \bibnamefont{and}
  \bibinfo{author}{\bibfnamefont{C.}~\bibnamefont{Wetterich}},
  \bibinfo{journal}{Phys. Rev.} \textbf{\bibinfo{volume}{D53}},
  \bibinfo{pages}{5142} (\bibinfo{year}{1996}), \eprint{hep-ph/9505267}.

\bibitem[{\citenamefont{Litim}(2000)}]{Litim:2000ci}
\bibinfo{author}{\bibfnamefont{D.~F.} \bibnamefont{Litim}},
  \bibinfo{journal}{Phys. Lett.} \textbf{\bibinfo{volume}{B486}},
  \bibinfo{pages}{92} (\bibinfo{year}{2000}), \eprint{hep-th/0005245}.

\bibitem[{\citenamefont{Litim}(2001{\natexlab{a}})}]{Litim:2001up}
\bibinfo{author}{\bibfnamefont{D.~F.} \bibnamefont{Litim}},
  \bibinfo{journal}{Phys. Rev.} \textbf{\bibinfo{volume}{D64}},
  \bibinfo{pages}{105007} (\bibinfo{year}{2001}{\natexlab{a}}),
  \eprint{hep-th/0103195}.

\bibitem[{\citenamefont{Litim}(2001{\natexlab{b}})}]{Litim:2001fd}
\bibinfo{author}{\bibfnamefont{D.~F.} \bibnamefont{Litim}},
  \bibinfo{journal}{Int. J. Mod. Phys.} \textbf{\bibinfo{volume}{A16}},
  \bibinfo{pages}{2081} (\bibinfo{year}{2001}{\natexlab{b}}),
  \eprint{hep-th/0104221}.

\bibitem[{\citenamefont{Reuter and Wetterich}(1994)}]{Reuter:1993kw}
\bibinfo{author}{\bibfnamefont{M.}~\bibnamefont{Reuter}} \bibnamefont{and}
  \bibinfo{author}{\bibfnamefont{C.}~\bibnamefont{Wetterich}},
  \bibinfo{journal}{Nucl. Phys.} \textbf{\bibinfo{volume}{B417}},
  \bibinfo{pages}{181} (\bibinfo{year}{1994}).

\bibitem[{\citenamefont{Gies}(2002)}]{Gies:2002af}
\bibinfo{author}{\bibfnamefont{H.}~\bibnamefont{Gies}}, \bibinfo{journal}{Phys.
  Rev.} \textbf{\bibinfo{volume}{D66}}, \bibinfo{pages}{025006}
  (\bibinfo{year}{2002}), \eprint{hep-th/0202207}.

\bibitem[{\citenamefont{Pawlowski et~al.}(2004)\citenamefont{Pawlowski, Litim,
  Nedelko, and von Smekal}}]{Pawlowski:2003hq}
\bibinfo{author}{\bibfnamefont{J.~M.} \bibnamefont{Pawlowski}},
  \bibinfo{author}{\bibfnamefont{D.~F.} \bibnamefont{Litim}},
  \bibinfo{author}{\bibfnamefont{S.}~\bibnamefont{Nedelko}}, \bibnamefont{and}
  \bibinfo{author}{\bibfnamefont{L.}~\bibnamefont{von Smekal}},
  \bibinfo{journal}{Phys. Rev. Lett.} \textbf{\bibinfo{volume}{93}},
  \bibinfo{pages}{152002} (\bibinfo{year}{2004}), \eprint{hep-th/0312324}.

\bibitem[{\citenamefont{van Ritbergen et~al.}(1997)\citenamefont{van Ritbergen,
  Vermaseren, and Larin}}]{vanRitbergen:1997va}
\bibinfo{author}{\bibfnamefont{T.}~\bibnamefont{van Ritbergen}},
  \bibinfo{author}{\bibfnamefont{J.~A.~M.} \bibnamefont{Vermaseren}},
  \bibnamefont{and} \bibinfo{author}{\bibfnamefont{S.~A.} \bibnamefont{Larin}},
  \bibinfo{journal}{Phys. Lett.} \textbf{\bibinfo{volume}{B400}},
  \bibinfo{pages}{379} (\bibinfo{year}{1997}), \eprint{hep-ph/9701390}.

\bibitem[{\citenamefont{Czakon}(2005)}]{Czakon:2004bu}
\bibinfo{author}{\bibfnamefont{M.}~\bibnamefont{Czakon}},
  \bibinfo{journal}{Nucl. Phys.} \textbf{\bibinfo{volume}{B710}},
  \bibinfo{pages}{485} (\bibinfo{year}{2005}), \eprint{hep-ph/0411261}.

\bibitem[{\citenamefont{Ellwanger}(1994)}]{Ellwanger:1994iz}
\bibinfo{author}{\bibfnamefont{U.}~\bibnamefont{Ellwanger}},
  \bibinfo{journal}{Phys. Lett.} \textbf{\bibinfo{volume}{B335}},
  \bibinfo{pages}{364} (\bibinfo{year}{1994}), \eprint{hep-th/9402077}.

\bibitem[{\citenamefont{von Smekal et~al.}(1997)\citenamefont{von Smekal,
  Alkofer, and Hauck}}]{vonSmekal:1997is}
\bibinfo{author}{\bibfnamefont{L.}~\bibnamefont{von Smekal}},
  \bibinfo{author}{\bibfnamefont{R.}~\bibnamefont{Alkofer}}, \bibnamefont{and}
  \bibinfo{author}{\bibfnamefont{A.}~\bibnamefont{Hauck}},
  \bibinfo{journal}{Phys. Rev. Lett.} \textbf{\bibinfo{volume}{79}},
  \bibinfo{pages}{3591} (\bibinfo{year}{1997}), \eprint{hep-ph/9705242}.

\bibitem[{\citenamefont{von Smekal et~al.}(1998)\citenamefont{von Smekal,
  Hauck, and Alkofer}}]{vonSmekal:1997vx}
\bibinfo{author}{\bibfnamefont{L.}~\bibnamefont{von Smekal}},
  \bibinfo{author}{\bibfnamefont{A.}~\bibnamefont{Hauck}}, \bibnamefont{and}
  \bibinfo{author}{\bibfnamefont{R.}~\bibnamefont{Alkofer}},
  \bibinfo{journal}{Ann. Phys.} \textbf{\bibinfo{volume}{267}},
  \bibinfo{pages}{1} (\bibinfo{year}{1998}), \eprint{hep-ph/9707327}.

\bibitem[{\citenamefont{Lerche and von Smekal}(2002)}]{Lerche:2002ep}
\bibinfo{author}{\bibfnamefont{C.}~\bibnamefont{Lerche}} \bibnamefont{and}
  \bibinfo{author}{\bibfnamefont{L.}~\bibnamefont{von Smekal}},
  \bibinfo{journal}{Phys. Rev.} \textbf{\bibinfo{volume}{D65}},
  \bibinfo{pages}{125006} (\bibinfo{year}{2002}), \eprint{hep-ph/0202194}.

\bibitem[{\citenamefont{Alkofer et~al.}(2005)\citenamefont{Alkofer, Fischer,
  and Llanes-Estrada}}]{Alkofer:2004it}
\bibinfo{author}{\bibfnamefont{R.}~\bibnamefont{Alkofer}},
  \bibinfo{author}{\bibfnamefont{C.~S.} \bibnamefont{Fischer}},
  \bibnamefont{and} \bibinfo{author}{\bibfnamefont{F.~J.}
  \bibnamefont{Llanes-Estrada}}, \bibinfo{journal}{Phys. Lett.}
  \textbf{\bibinfo{volume}{B611}}, \bibinfo{pages}{279} (\bibinfo{year}{2005}),
  \eprint{hep-th/0412330}.

\bibitem[{\citenamefont{Fischer et~al.}(2009)\citenamefont{Fischer, Maas, and
  Pawlowski}}]{Fischer:2008uz}
\bibinfo{author}{\bibfnamefont{C.~S.} \bibnamefont{Fischer}},
  \bibinfo{author}{\bibfnamefont{A.}~\bibnamefont{Maas}}, \bibnamefont{and}
  \bibinfo{author}{\bibfnamefont{J.~M.} \bibnamefont{Pawlowski}},
  \bibinfo{journal}{Annals Phys.} \textbf{\bibinfo{volume}{324}},
  \bibinfo{pages}{2408} (\bibinfo{year}{2009}), \eprint{0810.1987}.

\bibitem[{\citenamefont{Fischer and Pawlowski}(2009)}]{Fischer:2009tn}
\bibinfo{author}{\bibfnamefont{C.~S.} \bibnamefont{Fischer}} \bibnamefont{and}
  \bibinfo{author}{\bibfnamefont{J.~M.} \bibnamefont{Pawlowski}},
  \bibinfo{journal}{Phys. Rev.} \textbf{\bibinfo{volume}{D80}},
  \bibinfo{pages}{025023} (\bibinfo{year}{2009}), \eprint{0903.2193}.

\bibitem[{\citenamefont{Fischer and Pawlowski}(2007)}]{Fischer:2006vf}
\bibinfo{author}{\bibfnamefont{C.~S.} \bibnamefont{Fischer}} \bibnamefont{and}
  \bibinfo{author}{\bibfnamefont{J.~M.} \bibnamefont{Pawlowski}},
  \bibinfo{journal}{Phys. Rev.} \textbf{\bibinfo{volume}{D75}},
  \bibinfo{pages}{025012} (\bibinfo{year}{2007}), \eprint{hep-th/0609009}.

\bibitem[{\citenamefont{Aguilar et~al.}(2009)\citenamefont{Aguilar, Binosi,
  Papavassiliou, and Rodriguez-Quintero}}]{Aguilar:2009nf}
\bibinfo{author}{\bibfnamefont{A.~C.} \bibnamefont{Aguilar}},
  \bibinfo{author}{\bibfnamefont{D.}~\bibnamefont{Binosi}},
  \bibinfo{author}{\bibfnamefont{J.}~\bibnamefont{Papavassiliou}},
  \bibnamefont{and}
  \bibinfo{author}{\bibfnamefont{J.}~\bibnamefont{Rodriguez-Quintero}},
  \bibinfo{journal}{Phys. Rev.} \textbf{\bibinfo{volume}{D80}},
  \bibinfo{pages}{085018} (\bibinfo{year}{2009}), \eprint{0906.2633}.

\end{thebibliography}

\end{document}